\documentclass[10pt,a4paper,reqno]{amsart}
\usepackage{acronym}
\usepackage{amsfonts}
\usepackage{amsmath}
\usepackage{amssymb}
\usepackage{amsthm}
\usepackage{bm}
\usepackage{braket}
\usepackage{cite}
\usepackage{color}
\usepackage{geometry}
\usepackage{graphicx}
\usepackage{hyperref}
\usepackage{mathrsfs}
\usepackage{mathtools}
\usepackage{setspace}
\usepackage{stmaryrd}
\usepackage{subcaption}
\usepackage{tikz}

\usepackage{epsfig}
\usepackage{setspace}
\usepackage{booktabs}
\usepackage{threeparttable}
\usepackage{diagbox}

\newgeometry{top=2cm,bottom=2cm,outer=1.5cm,inner=1.5cm}
\newcommand{\bse}{\begin{subequations}}
\newcommand{\ese}{\end{subequations}}

\numberwithin{equation}{section}

\title[Solving localized wave solutions of the DNLS using an improved PINN method]{Solving localized wave solutions of the derivative nonlinear Schr\"{o}dinger equation using an improved PINN method}
\author{Juncai Pu}
\address[JP]{School of Mathematical Sciences, Shanghai Key Laboratory of Pure Mathematics and Mathematical Practice, and Shanghai Key Laboratory of Trustworthy Computing \\
East China Normal University \\ Shanghai 200241 \\ People's Republic of China}
\author{Jun Li}
\address[JL]{Shanghai Key Laboratory of Trustworthy Computing \\ East China Normal University \\ Shanghai 200062\\
People's Republic of China}
\author{Yong Chen$^*$}
\address[YC]{School of Mathematical Sciences, Shanghai Key Laboratory of Pure Mathematics and Mathematical Practice, and Shanghai Key Laboratory of Trustworthy Computing \\
East China Normal University \\ Shanghai 200241 \\ People's Republic of China}
\address[YC]{College of Mathematics and Systems Science \\ Shandong University of Science and Technology \\ Qingdao 266590 \\ People's Republic of China}
\address[YC]{Department of Physics \\ Zhejiang Normal University \\ Jinhua 321004 \\ People's Republic of China}
\email{ychen@sei.ecnu.edu.cn}

\begin{document}

\begin{abstract}
The solving of the derivative nonlinear Schr\"odinger equation (DNLS) has attracted considerable attention in theoretical analysis and physical applications. Based on the physics-informed neural network (PINN) which has been put forward to uncover dynamical behaviors of nonlinear partial different equation from spatiotemporal data directly, an improved PINN method with neuron-wise locally adaptive activation function is presented to derive localized wave solutions of the DNLS in complex space. In order to compare the performance of above two methods, we reveal the dynamical behaviors and error analysis for localized wave solutions which include one-rational soliton solution, genuine rational soliton solutions and rogue wave solution of the DNLS by employing two methods, also exhibit vivid diagrams and detailed analysis. The numerical results demonstrate the improved method has faster convergence and better simulation effect. On the bases of the improved method, the effects for different numbers of initial points sampled, residual collocation points sampled, network layers, neurons per hidden layer on the second order genuine rational soliton solution dynamics of the DNLS are considered, and the relevant analysis when the locally adaptive activation function chooses different initial values of scalable parameters are also exhibited in the simulation of the two-order rogue wave solution.
\end{abstract}

\maketitle

\section{Introduction}

The derivative nonlinear Schr\"{o}dinger equation (DNLS)
\begin{equation}\label{e1}
iq_t+q_{xx} + i(q^2q^*)_x=0,
\end{equation}
plays a significant role both in the integrable system theory and many physical applications, especially in space plasma physics and nonlinear optics \cite{Kaup1978, Mjolhus1976}. Here, $q=q(x,t)$ are complex-valued solutions, the superscript $``*"$ denotes complex conjugation, and the subscripts $x$ and $t$ denote the partial derivatives with respect to $x$ and $t$, respectively. In recent decades, many scholars have invested a lot of time and energy to study various mathematical and physical problems of the DNLS. Mio et al. derived the DNLS of Alfven waves in plasma, and it well describes the propagation of small amplitude nonlinear Alfven waves in low-$\beta$ plasma, propagating strictly parallel or at a small angle to the ambient magnetic field \cite{Mio1976, Huang1990}. The results show that the large amplitude Magneto-Hydro-Dynamical waves propagating at arbitrary angle with the surrounding magnetic field in high $\beta$ plasma are also simulated by the DNLS. In nonlinear optics, the modified nonlinear Schr\"{o}dinger equation, which is gauge equivalent to the DNLS, is derived in the theory of ultrashort femtosecond nonlinear pulse in optical fiber \cite{ChenX2004}. While the spectrum width of the pulse is equal to the carrier frequency, the self steepening effect of the pulse should be considered. In addition, the filamentation of lower-hybrid waves can be simulated by the DNLS which governs the asymptotic state of the filamentation, and it admits moving solitary envelope solutions for the electric field \cite{Spatchek1977}. Ichikawa and co-workers obtained the peculiar structure of spiky modulation of amplitude and phase, which is arisen from the derivative nonlinear coupling term \cite{Ichikawa1980}. At present, the abundant solutions and integrability of the DNLS have been derived through different methods. Kaup and Newell proved the integrability of the DNLS in the sense of inverse scattering method in 1978 \cite{Kaup1978}. Nakamura and Chen constructed the first N-soliton formula of the DNLS with the help of the Hirota bilinear transformation method \cite{Nakamura1980}. Furthermore, based on Darboux transform technique, Huang and Chen established the determinant form of N-soliton formula \cite{Huang1990}. Kamchatnov and cooperators not only proposed a method for finding periodic solutions of several integrable evolution equations and applied it to the DNLS, but also dealt with the formation of solitons on the sharp front of optical pulse in an optical fiber according to the DNLS \cite{Kamchatnov1990,Kamchatnov1998}. The Cauchy problem of the DNLS has been discussed by Hayashi and Ozawa \cite{Hayashi1992}. The compact N-soliton formulae both with asymptotically vanishing and non-vanishing amplitudes were obtained by iterating B\"acklund transformation of the DNLS \cite{Steudel2003}. In addition, the high-order solitons, high-order rogue waves, and rational solutions for the DNLS have been given out explicitly with the help of two kinds of generalized Darboux transformations which rely on certain limit technique \cite{Guo2012}. Recently, more abundant solutions and new physical phenomena of the DNLS are revealed by various methods \cite{XuT2018,Xue2020,Xu2019,Zhang2020,Wang2015,Yang2020}.

In recent years, due to the explosive growth of available data and computing resources, neural networks(NNs) have been successfully applied in diverse fields, such as recommendation system, speech recognition, mathematical physics, computer vision, pattern recognition and so on \cite{LeCun2015,Bishop2006,Li2020,Krizhevsky2017,Lake2015}. Particularly, a physics-informed neural network method (PINN) has been proved to be particularly suitable for solving and inversing equations which have been controlled mathematical physical systems on the basis of NNs, and found that the high-dimensional network tasks can be completed with less data sets\cite{Raissi2019,Lij2020}. The PINN method can not only accurately solve both forward problems, where the approximate solutions of governing equations are obtained, but also precisely deal with the highly ill-posed inverse problems, where parameters involved in the governing equation are inferred from the training data. Based on the abundant solutions and integrability of the integrable systems \cite{Pu2020,Hirota2004,Zakharov1984}, we have simulated the one and two order rogue wave solutions of the integrable nonlinear Schr\"odinger equation by employing the deep learning method with physical constraints \cite{PuJ2020}. The slow convergence performance leads to the increase of training time and higher performance requirements of experimental equipment, so it is essential to accelerate the convergence of the network without sacrificing the performance. Meanwhile, the original PINN method could not accurately reconstruct the complex solutions in some complicated equations. it is crucial to design a higher efficient and more adaptable deep learning algorithm to not only improve the accuracy of the simulated solution but also reduce the training cost.

As is known to all, a significant feature of NNs is the activation function, which determines the activation of specific neurons and the stability of network performance in the training process. There is just a rule-of-thumb for the choice of activation function, which depends entirely on the problem at hand. In the PINN algorithm, many activation functions such as the sigmoid function, tanh, sin etc are used to solve various problems, refer to \cite{LiJ2020,Raissi2019} for details. Recently, a variety of research methods for activation functions have been proposed to optimize convergence performance and raise the training speed. Dushkoff and Ptucha proposed multiple activation functions of per neuron, in which individual neuron chooses between multiple activation functions \cite{Dushkoff2016}. Li et al. proposed a tunable activation function while only one hidden layer is used \cite{LiB2013}. The authors focused on learning activation functions in convolutional NNs by combining basic activation functions in a data-driven way \cite{Qian2018}. Jagtap and collaborators employed adaptive activation functions for regression in PINN to approximate smooth and discontinuous functions as well as solutions of linear and nonlinear partial differential equations, and introduced a scalable parameters in the activation function, which can be optimized to achieve best performance of the network as it changes dynamically the topology of the loss function involved in the optimization process \cite{Jagtap2020}. The adaptive activation function has better learning capabilities than the traditional fixed activation as it improves greatly the convergence rate, especially during early training, as well as the solution accuracy.

In particular, Jagtap et al. presented two different kinds of locally adaptive activation functions, namely layer-wise and neuron-wise locally adaptive activation functions \cite{JagtapA2020}. Compared with global adaptive activation functions \cite{Jagtap2020}, the locally adaptive activation functions further improve the training speed and performance of NNs. Furthermore, in order to further speed up the training process, a slope recovery term based on activation slope has been added to the loss function of layer-wise and neuron-wise locally adaptive activation functions to improve the performance of neural network. Recently, we focus on studying abundant solutions of integrable equations \cite{Li2020,Lij2020,LiJ2020,PuJ2020} due to they have better integrability such as Painlev\'{e} integrability, Lax integrability, Liouville integrability and so on \cite{Weiss1983,Lax1968,Tu1989}. Significantly, the DNLS has been pointed out that it satisfy important integrability properties, and many types of localized wave solutions have been obtained by various effective methods \cite{Kaup1978,Mjolhus1976,Mio1976,Huang1990,ChenX2004}. We extend the PINN based on locally adaptive activation function with slope recovery term which proposed by Jagtap and cooperator \cite{JagtapA2020} to solve the nonlinear integrable equation in complex space, and construct the localized wave solutions which consist of the rational soliton solutions and rogue wave solution of the integrable DNLS. Meanwhile, we also demonstrate the relevant results that contain the rational soliton solutions and rogue wave solution by exploiting the PINN, which are convenient for comparative analysis. The performance comparison between the improved PINN method with locally adaptive activation functions and the PINN method are given out in detail.

This paper is organized as follows. In section 2, we introduce briefly discussions of the original PINN method and improved PINN method with locally adaptive activation function, where also discuss about training data, loss function, optimization method and the operating environment. In Section 3, the one-rational soliton solution and the first order genuine rational soliton solution of the DNLS are obtained by two distinct PINN approaches. Section 4 provides the second order genuine rational solution and two-order rogue wave solution for the DNLS, and the relative $\mathbb{L}_2$ errors of simulating the second order genuine rational solution of the DNLS with different numbers of initial points sampled, residual collocation points sampled, network layers and neurons per hidden layer are also given out in detail. Moreover, the effects of the initial values of scalable parameters on the two-order rogue wave solution are shown. Conclusion is given out in last section.

\section{Methodology}
Here, we will consider the general (1+1)-dimensional nonlinear time-dependent integrable equations in complex space, where each contains a dissipative term as well as other partial derivatives, such as nonlinear terms or dispersive terms, as follows
\begin{equation}\label{e2}
q_t+\mathcal{N}(q,q_x,q_{xx},q_{xxx},\cdots)=0,
\end{equation}
where $q$ are complex-valued solutions of $x$ and $t$ to be determined later, and $\mathcal{N}$ is a nonlinear functional of the solution $q$ and its derivatives of arbitrary orders with respect to $x$. Due to the complexity of the structure of the solution $q(x,t)$ of Eq. \eqref{e2}, we decompose $q(x,t)$ into the real part $u(x,t)$ and the imaginary part $v(x,t)$, i.e. $q=u+iv$. It is obvious that $u(x,t)$ and $v(x,t)$ are real-valued functions. Then substituting into Eq. \eqref{e2}, we have
\begin{equation}\label{e3}
u_t+\mathcal{N}_u(u,u_x,u_{xx},u_{xxx},\cdots)=0,
\end{equation}
\begin{equation}\label{e4}
v_t+\mathcal{N}_v(v,v_x,v_{xx},v_{xxx},\cdots)=0,
\end{equation}
where $\mathcal{N}_u$ and $\mathcal{N}_v$ are nonlinear functionals of the corresponding solution and its derivatives of arbitrary orders with respect to $x$, respectively. In this section, we will briefly introduce the original PINN method and its improved version, respectively.

\subsection{The PINN method}
\quad

Here, we first construct a simple multilayer feedforward neural network with depth $D$ which contains an input layer, $D-1$ hidden layers and an output layer. Without loss of generality, we assume that there are $N_d$ neurons in the $d^{th}$ hidden layer. Then, the $d^{th}$ hidden layer receives the post-activation output $\textbf{x}^{d-1}\in \mathbb{R}^{N_{d-1}}$ of the previous layer as its input, and the specific affine transformation is of the form
\begin{equation}\label{e5}
\mathcal{L}_d\left(\textbf{x}^{d-1}\right)\triangleq \textbf{W}^{d}\textbf{x}^{d-1}+\textbf{b}^{d},
\end{equation}
where the network weight $\textbf{W}^d\in \mathbb{R}^{N_d\times N_{d-1}}$ and the bias term $\textbf{b}^d\in \mathbb{R}^{N_d}$ to be learned are initialized using some special strategies, such as Xavier initialization or He initialization \cite{Glorot2010,He2015}.

The nonlinear activation function $\sigma(\cdot)$ is applied component-wise to the affine output $\mathcal{L}_d$ of the present layer. In addition, this nonlinear activation is not applied in the output layer for some regression problems, or equivalently, we can say that the identity activation is used in the output layer. Therefore, the neural network can be represented as
\begin{equation}\label{e6}
q(\textbf{x};\Theta)=\left(\mathcal{L}_D\circ\sigma\circ\mathcal{L}_{D-1}\circ\cdots\circ\sigma\circ\mathcal{L}_1\right)(\textbf{x}),
\end{equation}
where the operator $``\circ"$ is the composition operator, $\Theta=\left\{\textbf{W}^d,\textbf{b}^d\right\}_{d=1}^{D}\in \mathcal{P}$ represents the learnable parameters to be optimized later in the network, and $\mathcal{P}$ is the parameter space, $q$ and $\textbf{x}^0=\textbf{x}$ are the output and input of the network, respectively.

The universal approximation property of the neural network and the idea of physical constraints play key roles in the PINN method. Thus, based on the PINN method \cite{Raissi2019}, we can approximate the potential complex-valued solution $q(x,t)$ of nonlinear integrable equations using a neural network. Then, the underlying laws of physics described by the governing equations are embedded into the network. By the aid of automatic differentiation (AD) mechanism in deep learning \cite{Baydin2018}, we can automatically and conveniently obtain the derivatives of the solution with respect to its inputs, i.e., the time and space coordinates. Compared with the traditional numerical differentiation methods, AD is a mesh-free method and does not suffer from some common errors, such as the truncation errors and round-off errors. To a certain extent, this AD technique enables us to open the black box of the neural network. In addition, the physics constraints can be regarded as a regularization mechanism that allows us to accurately recover the solution using a relatively simple feedforward network and remarkably few amounts of data. Moreover, the underlying physical laws introduce part interpretability into the neural network.

Specifically, we define the residual networks $f_u(x, t)$ and $f_v(x, t)$, which are given by the left-hand-side of Eq. \eqref{e3} and \eqref{e4}, respectively
\begin{equation}\label{e7}
f_u :=u_t+\mathcal{N}_u(u,u_x,u_{xx},u_{xxx},\cdots),
\end{equation}
\begin{equation}\label{e8}
f_v :=v_t+\mathcal{N}_v(v,v_x,v_{xx},v_{xxx},\cdots).
\end{equation}

Then the solution $q(x,t)$ will be trained to satisfy these two physical constraint conditions \eqref{e7} and \eqref{e8}, which play a vital role of regularization and have been embedded into the mean-squared objective function, that is, the loss function
\begin{equation}\label{e9}
Loss_{\Theta}=Loss_u+Loss_v+Loss_{f_u}+Loss_{f_v},
\end{equation}
where
\begin{equation}\label{e10}
Loss_u=\frac{1}{N_q}\sum^{N_q}_{i=1}|u(x_u^i,t_u^i)-u^i|^2,Loss_v=\frac{1}{N_q}\sum^{N_q}_{i=1}|v(x_v^i,t_v^i)-v^i|^2,
\end{equation}
and
\begin{equation}\label{e11}
Loss_{f_u}=\frac{1}{N_f}\sum^{N_f}_{j=1}|f_u(x_{f_u}^j,t_{f_u}^j)|^2,Loss_{f_v}=\frac{1}{N_f}\sum^{N_f}_{j=1}|f_v(x_{f_v}^j,t_{f_v}^j)|^2.
\end{equation}

Here $\{x^i_u,t^i_u,u^i\}^{N_q}_{i=1}$ and $\{x^i_v,t^i_v,v^i\}^{N_q}_{i=1}$ denote the initial and boundary value data of $q(x,t)$. Similarly, the collocation points for $f_u(x,t)$ and $f_v(x,t)$ are specified by $\{x_{f_u}^j,t_{f_u}^j\}^{N_{f}}_{j=1}$ and $\{x_{f_v}^j,t_{f_v}^j\}^{N_{f}}_{j=1}$. The loss function \eqref{e9} consists of the initial-boundary value data and the structure imposed by Eq. \eqref{e7} and \eqref{e8} at a finite set of collocation points. Specifically, the first and second terms on the right hand side of Eq. \eqref{e9} attempt to fit the solution data, and the third and fourth terms on the right hand side learn to discover the real solution space.

\subsection{The improved PINN method}
\quad

The original PINN method could not accurately reconstruct complex solutions in some complicated nonlinear integrable equations. Thus, we present an improved PINN method (IPINN) where a locally adaptive activation function technique is introduced into the original PINN method. It changes the slope of the activation function adaptively, resulting in non-vanishing gradients and faster training of the network. There are several kinds of locally adaptive activation functions, for example, layer-wise and neuron-wise. In this paper, we only consider the neuron-wise version due to some accuracy and performance requirements. Specifically, we first define such activation function as
\begin{equation}\nonumber
\sigma\left(na^d_i\left(\mathcal{L}_d\left(\textbf{x}^{d-1}\right)\right)_i\right),d=1,2,\cdots,D-1,i=1,2,\cdots,N_d,
\end{equation}
where $n>1$ is a scaling factor and $\{a^d_i\}$ are additional $\Sigma_{d=1}^{D-1}N_d$ parameters to be optimized. Note that, there is a critical scaling factor $n_{c}$ above which the optimization algorithm will become sensitive in each problem set. The neuron activation function acts as a vector activation function in each hidden layer, and each neuron has its own slope of activation function.

Based on Eq. \eqref{e6}, the new neural network with neuron-wise locally adaptive activation function can be represented as
\begin{equation}\label{e12}
q(\textbf{x};\bar{\Theta})=\left(\mathcal{L}_D\circ\sigma\circ na^{D-1}_{i}\left(\mathcal{L}_{D-1}\right)_{i}\circ\cdots\circ\sigma\circ na^1_i\left(\mathcal{L}_1\right)_i\right)(\textbf{x}),
\end{equation}
where the set of trainable parameters $\bar{\Theta}\in\bar{\mathcal{P}}$ consists of $\left\{\textbf{W}^d,\textbf{b}^d\right\}_{d=1}^{D}$ and $\left\{a_i^d\right\}_{i=1}^{D-1},\forall i=1,2,\cdots,N_d$, $\bar{\mathcal{P}}$ is the parameter space. In this method, the initialization of scalable parameters are carried out in the case of $na_i^d=1,\forall n\geqslant1$.

The resulting optimization algorithm will attempt to find the optimized parameters including the weights, biases, and additional coefficients in the activation to minimize the new loss function defined as
\begin{equation}\label{e13}
Loss_{\bar{\Theta}}=Loss_u+Loss_v+Loss_{f_u}+Loss_{f_v}+Loss_S,
\end{equation}
where $Loss_u, Loss_v, Loss_{f_u}$ and $Loss_{f_v}$ are defined by Eqs. \eqref{e10}-\eqref{e11}. The last slope recovery term $Loss_S$ in the loss function \eqref{e13} is defined as
\begin{equation}\label{e14}
Loss_S=\frac{1}{\frac{1}{D-1}\sum_{d=1}^{D-1}\mathrm{exp}\left(\frac{\sum_{i=1}^{N_d}a_i^d}{N_d}\right)},
\end{equation}

This term $Loss_S$ forces the neural network to increase the activation slope value quickly, which ensures the non-vanishing of the gradient of the loss function and improves the network's training speed. Compared with the PINN method in Section 2.1, the improved method induces a new gradient dynamics, which results in better convergence points and faster convergence rate. Jagtap et al. stated that a gradient descent algorithm such as stochastic gradient descent (SGD) minimizing the loss function \eqref{e13} does not converge to a sub-optimal critical point or a sub-optimal local minimum, for the neuron-wise locally adaptive activation function, given certain appropriate initialization and learning rates \cite{JagtapA2020}.

In both methods, all loss functions are simply optimized by employing the L-BFGS algorithm, which is a full-batch gradient descent optimization algorithm based on a quasi-Newton method \cite{Liu1989}. Especially, the scalable parameters in the adaptive activation function are initialized generally as $n=10,a_i^d=0.1$, unless otherwise specified. In addition, we select relatively simple multi-layer perceptrons (i.e., feedforward neural networks) with the Xavier initialization and the $\tanh$ activation function. All the codes in this article is based on Python 3.8 and Tensorflow 1.15, and all numerical experiments reported here are run on a DELL Precision 7920 Tower computer with 2.10 GHz 8-core Xeon Silver 4110 processor and 64 GB memory.

\section{One-rational soliton solution and first order genuine rational soliton solution of the DNLS}

In this section, two different neural network methods mentioned in the previous section are used to obtain the simulation solution of the DNLS, and the dynamic behavior, error analysis and related plots of the one-rational soliton solution and first order genuine rational soliton solution for the DNLS are listed out in detail. We consider the DNLS along with Dirichlet boundary conditions given by
\begin{equation}\label{e15}
\begin{split}
\begin{cases}
iq_t+q_{xx} + i(q^2q^*)_x=0,x\in[x_0,x_1],t\in[t_0,t_1],\\
q(x,t_0)=q_0(x),\\
q(x_0,t)=q(x_1,t),\\
\end{cases}
\end{split}
\end{equation}
where $x_0,x_1$ represent the lower and upper boundaries of $x$ respectively. Similarly, $t_0$ and $t_1$ represent the initial and final times of $t$ respectively. The initial condition $q_0(x)$ is an arbitrary complex-valued function. The rational soliton solutions of the DNLS have been obtained by generalized Darboux transformations \cite{Guo2012}. In this part, we will employ two different types of approaches which contain the PINN and IPINN to simulate two different forms of rational soliton solutions. Compared with the known exact solutions of the DNLS, so as to prove that the numerical solutions $q(x,t)$ obtained by neural network models is effective. From Ref. \cite{Guo2012}, we can derived the form about one-rational soliton solution and first order genuine rational soliton solution of the DNLS. the one-rational soliton solution formulation shown as follow
\begin{equation}\label{e16}
q(x,t)=\frac{4a^3[4i(a^2x-4t+a^2c)-a^4]e^{\frac{2i(a^2x-2t+a^2c)}{a^4}}}{[4i(a^2x-4t+a^2c)+a^4]^2},
\end{equation}
where $a,c$ are arbitrary constants, $i^2=-1$. Therefore, the velocity for this one-rational soliton solution is $a^2/4$ and the center is along the line $a^2x-4t+a^2c=0$, the altitude for $|q(x,t)|$ is $16/a^2$.

On the other hand, the first order genuine rational soliton solution of the DNLS can be represented as following
\begin{equation}\label{e17}
q(x,t)=-\frac{(-2x+6t-i)(-2x+6t+3i)}{(-2x+6t+i)^2},
\end{equation}
which is nothing but the rational traveling wave solution with non-vanishing background. In the next two sections, we use the PINN method and the improved PINN method to simulate the above two solutions, respectively. Some necessary comparisons and analyses are exhibited in detail.

\subsection{One-rational soliton solution}
\quad

In this section, based on the neural network structure which contains nine hidden layers, each layer has 40 neurons, we numerically construct one-rational soliton solution of the DNLS via the PINN method and improved PINN method. One can obtain the exact one-rational soliton solution of Eq. \eqref{e15} after taking $a=1,c=1$ into Eq. \eqref{e16} as follow
\begin{equation}\label{e18}
q(x,t)=\frac{4[4i(1-4t+x)-1]e^{2i(1-2t+x)}}{[4i(1-4t+x)+1]^2}.
\end{equation}

Then we take $[x_0,x_1]$ and $[t_0,t_1]$ in Eq. \eqref{e15} as $[-2.0,0.0]$ and $[-0.1,0.1]$, respectively. The corresponding initial condition $q_0(x)$ is obtained by substituting a specific initial value into \eqref{e18}
\begin{equation}\label{e19}
q_0(x)=\frac{4[4i(1.4+x)-1]e^{2i(1.2+x)}}{[4i(1.4+x)+1]^2}.
\end{equation}

We employ the traditional finite difference scheme on even grids in MATLAB to simulate Eq. \eqref{e15} with the initial data \eqref{e18} to acquire the training data. In particular, the initialization of scalable parameters is $n=5, a_i^m=0.2$. Specifically, divide space $[-2.0,0.0]$ into 513 points and time $[-0.1,0.1]$ into 401 points, one-rational soliton solution $q(x,t)$ is discretized into $401$ snapshots accordingly. We generate a smaller training dataset that containing initial-boundary data by randomly extracting $N_q=100$ from original dataset and $N_f=10000$ collocation points which are generated by the Latin hypercube sampling method \cite{Stein1987}. After giving a dataset of initial and boundary points, the latent one-rational soliton solution $q(x,t)$ has been successfully learned by tuning all learnable parameters of the neural network and regulating the loss function \eqref{e9} and \eqref{e13}. The model of PINN achieves a relative $\mathbb{L}_2$ error of 4.345103$\mathrm{e}-$02 in about $1314.1089$ seconds, and the number of iterations is 15395. Nevertheless, the network of IPINN achieves a relative $\mathbb{L}_2$ error of 1.998304$\mathrm{e}-$02 in about $1358.9031$ seconds, and the number of iterations is 10966.

In Figure 1 and Figure 2, the density plots, the sectional drawing of the latent one-rational soliton solution $q(x,t)$ and the iteration number curve plots under PINN and IPINN structures are plotted respectively. The pictures (a) in Fig. 1 and Fig. 2 clearly compare the exact solution and the predicted spatiotemporal solution of the two different methods, respectively. We particularly present a comparison between the exact solution and the predicted solution at different times $t = -0.05, 0, 0.05$ in the bottom panel of (a) in Fig. 1 and Fig. 2. Obviously, the bottom panel of picture (a) in Figure 2 shows that the predicted solution of DNLS equation is more consistent with the exact solution than the bottom panel of picture (a) in Figure 1. In other words, the simulation effect of IPINN is better than PINN. It is not hard to see that the training loss curve of picture (b) in Fig. 2, which revealing the relation between iteration number and loss function, is more smooth and stable than the curve of picture (b) in Fig. 1. In this test case, the IPINN with slope recovery term perform better than PINN in terms of convergence speed and accuracy of the solution.

\begin{figure}
\centering
\includegraphics[width=6.5cm,height=5cm]{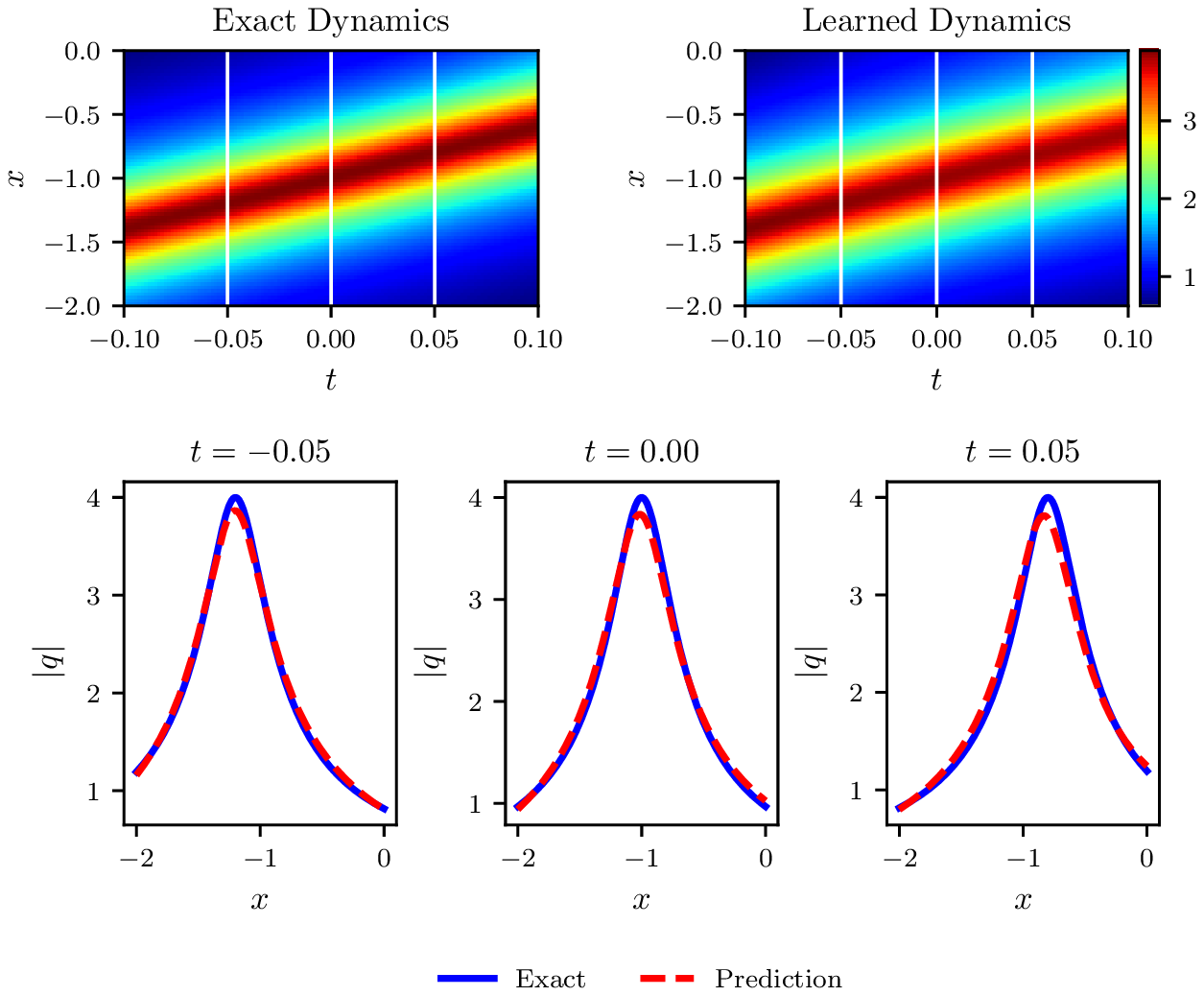}
$a$
\includegraphics[width=6.5cm,height=5cm]{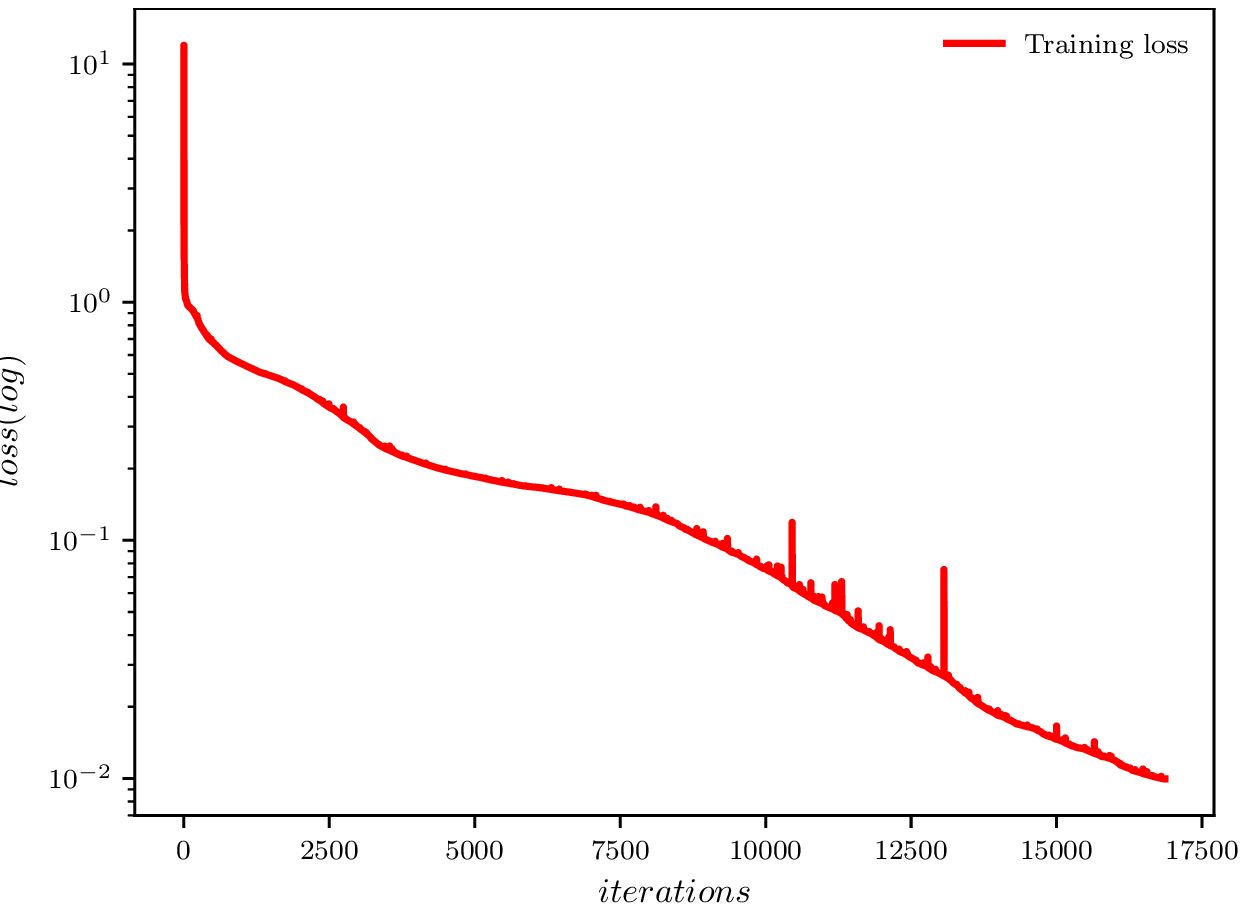}
$b$
\caption{The one-rational soliton solution $q(x,t)$ based on the PINN: (a) The density plots and the sectional drawing; (b) The loss curve figure.}
\end{figure}

\begin{figure}
\centering
\includegraphics[width=6.5cm,height=5cm]{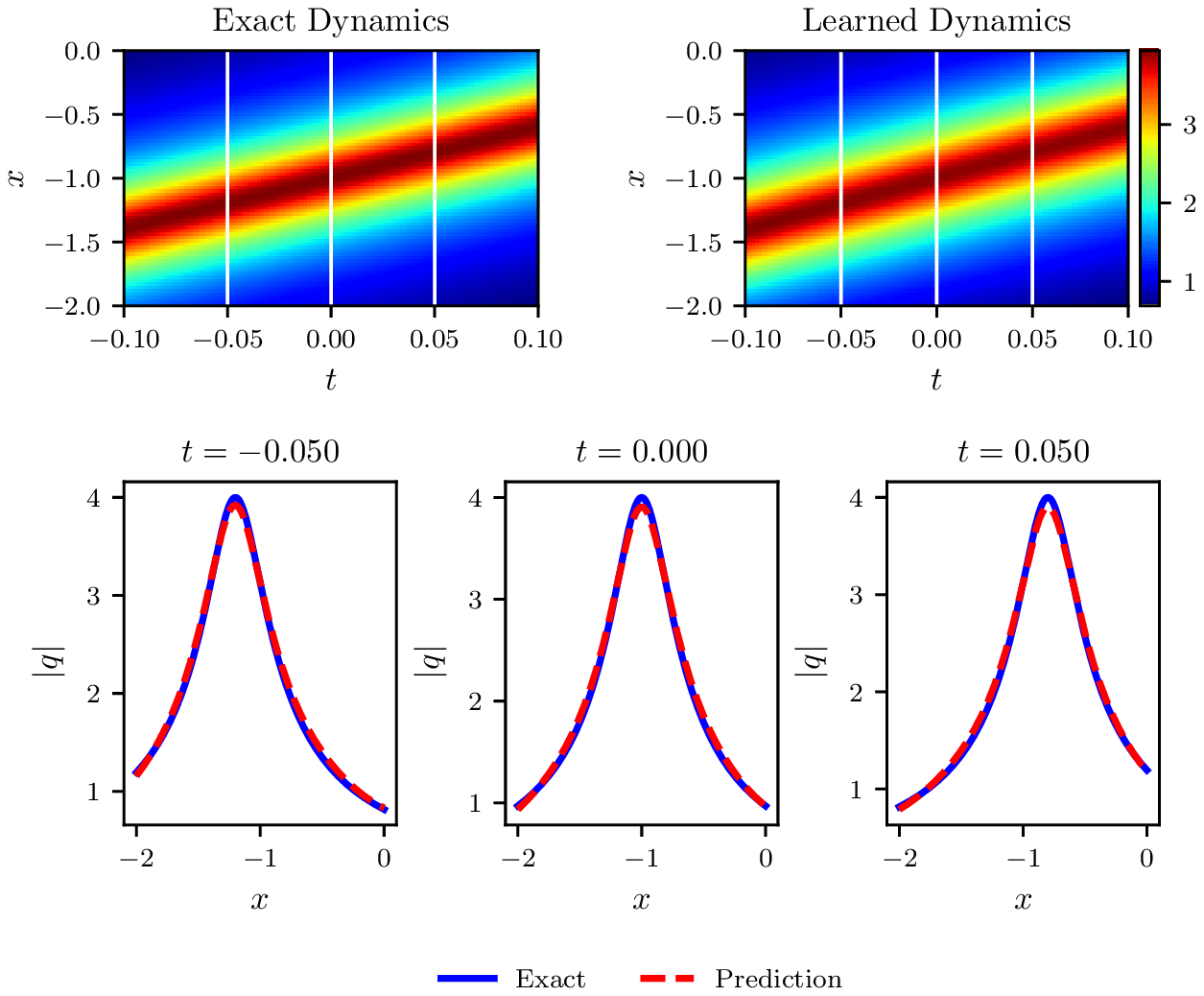}
$a$
\includegraphics[width=6.5cm,height=5cm]{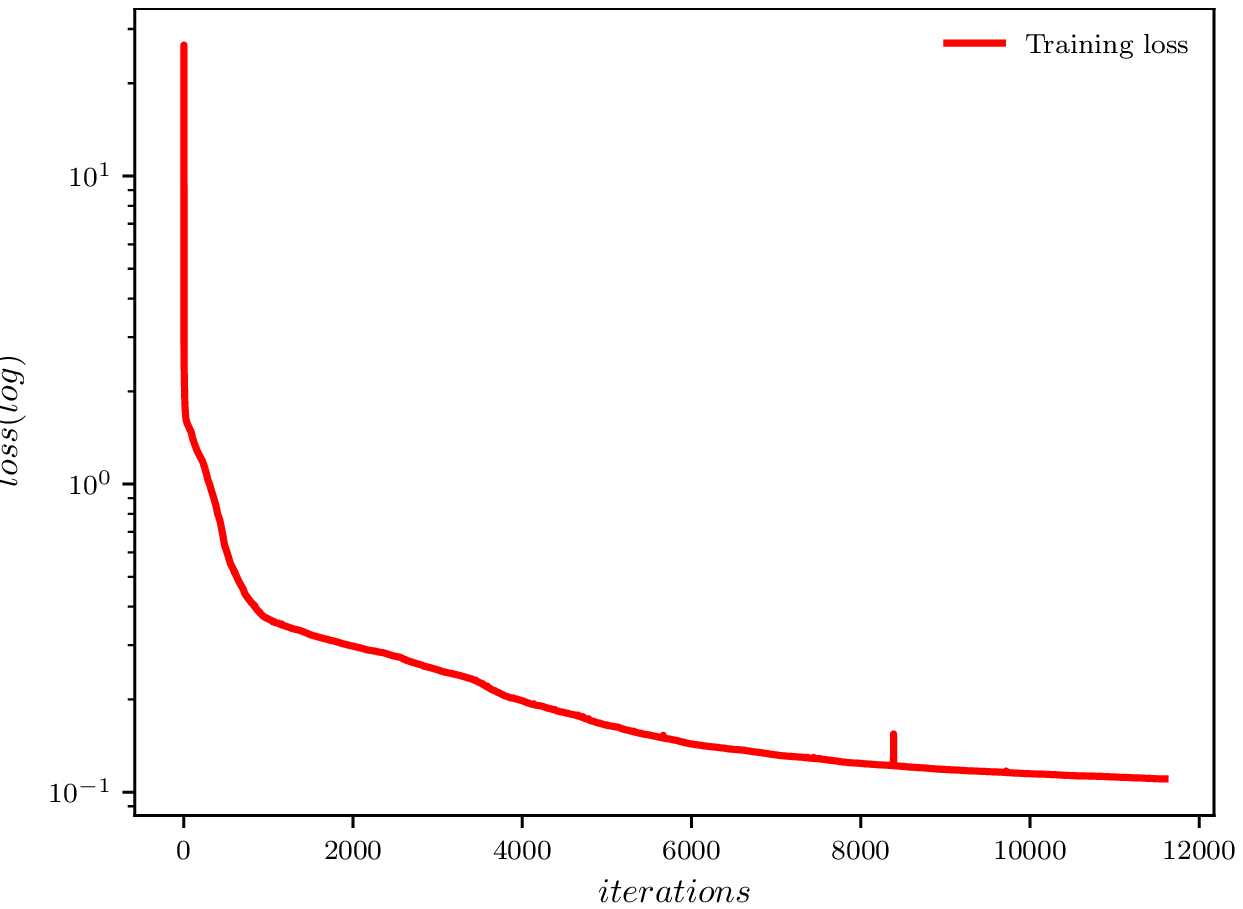}
$b$
\caption{The one-rational soliton solution $q(x,t)$ based on the IPINN: (a) The density plots and the sectional drawing; (b) The loss curve figure.}
\end{figure}

\subsection{First order genuine rational soliton solution}
\quad

In this section, we numerically construct the first order genuine rational soliton solution of Eq. \eqref{e15} by using the PINN and IPINN in which both contains nine hidden layers, each layer has 40 neurons. Now we take $[x_0,x_1]$ and $[t_0,t_1]$ in Eq. \eqref{e15} as $[-5.0,5.0]$ and $[-0.8,0.8]$, respectively. For instance, we consider the initial condition of first order genuine rational soliton solution based on Eq. \eqref{e17} is as follows
\begin{equation}\label{e20}
q_0(x)=-\frac{(-2x-4.8-i)(-2x-4.8+3i)}{(-2x-4.8+i)^2}.
\end{equation}

With the same data generation and sampling method in Section 3.1, and we numerically simulate the first order genuine rational soliton solution of the DNLS \eqref{e1} by using the PINN and IPINN, respectively. The training dataset that composed of initial-boundary data and collocation points is produced via randomly subsampling $N_q = 100$ from the original dataset and selecting $N_f = 10000$ configuration points which are generated by LHS. After training the first order genuine rational soliton solution with the help of PINN, the neural network achieves a relative $\mathbb{L}_2$ error of 5.598548$\mathrm{e}-$03 in about 349.5862 seconds, and the number of iterations is 3305. However, the network model by using the improved PINN method achieves a relative $\mathbb{L}_2$ error of 4.969464$\mathrm{e}-$03 in about 1103.1358 seconds, and the number of iterations is 10384. Apparently, when simulating the first order genuine rational soliton solution, the IPINN has more iterations, longer training time and smaller $\mathbb{L}_2$ error than the PINN.

Fig. 3 shows the density plots, profile and loss curve plots of the first order genuine rational soliton solution by employing the PINN. Figure 4 illustrates the density diagrams, profiles at different instants, error dynamics diagrams, three dimensional motion and loss curve figure of the first order genuine rational soliton solution based on the IPINN. We can clearly see that both methods can accurately simulate the first order genuine rational soliton solution From the $(a)$ in Fig. 3 and Fig. 4. However, comparing the b-graph of Fig. 3 with the d-graph of Fig. 4, we can clearly observe that the loss function curve of the IPINN decreases faster and smoother, while the loss function curve of PINN fluctuates greatly when the number of iterations is about 1500, and the burr phenomenon is remarkable obvious in the whole PINN training process. Furthermore, we can also gain that the ideal effect has been achieved when the IPINN is used for training after 2000 iterations from the loss function curve in Figure 4, so we can artificially control the appropriate number of iterations to save the training cost in some specific cases. At $t=-0.40,0,0.40$, we reveal the profiles of the three moments in bottom rows of $(a)$ in Fig. 3 and Fig. 4, and find the first order genuine rational solution has the property of soliton due to the amplitude does not change with time. The $(b)$ of Fig. 4 exhibt the error dynamics of the difference value between the exact solution and the predicted solution for the first order genuine rational soliton solution. In Fig. 4, the corresponding plot3d of the first order genuine rational soliton solution is plotted, it is evident that the first order genuine rational soliton solution is similar to the single-soliton solution with $|q|=1$ plane wave.

\begin{figure}
\centering
\includegraphics[width=6.5cm,height=5cm]{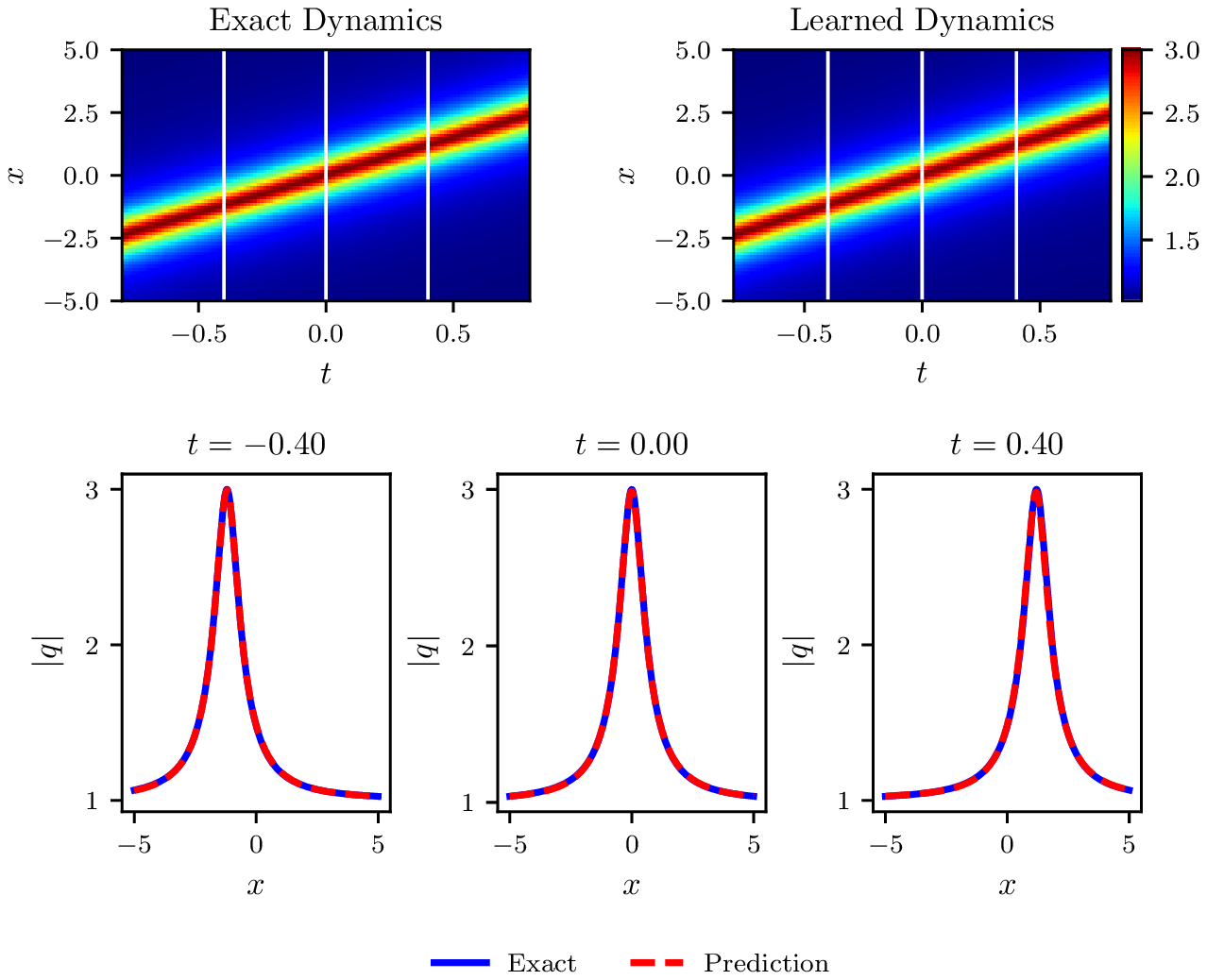}
$a$
\includegraphics[width=6.5cm,height=5cm]{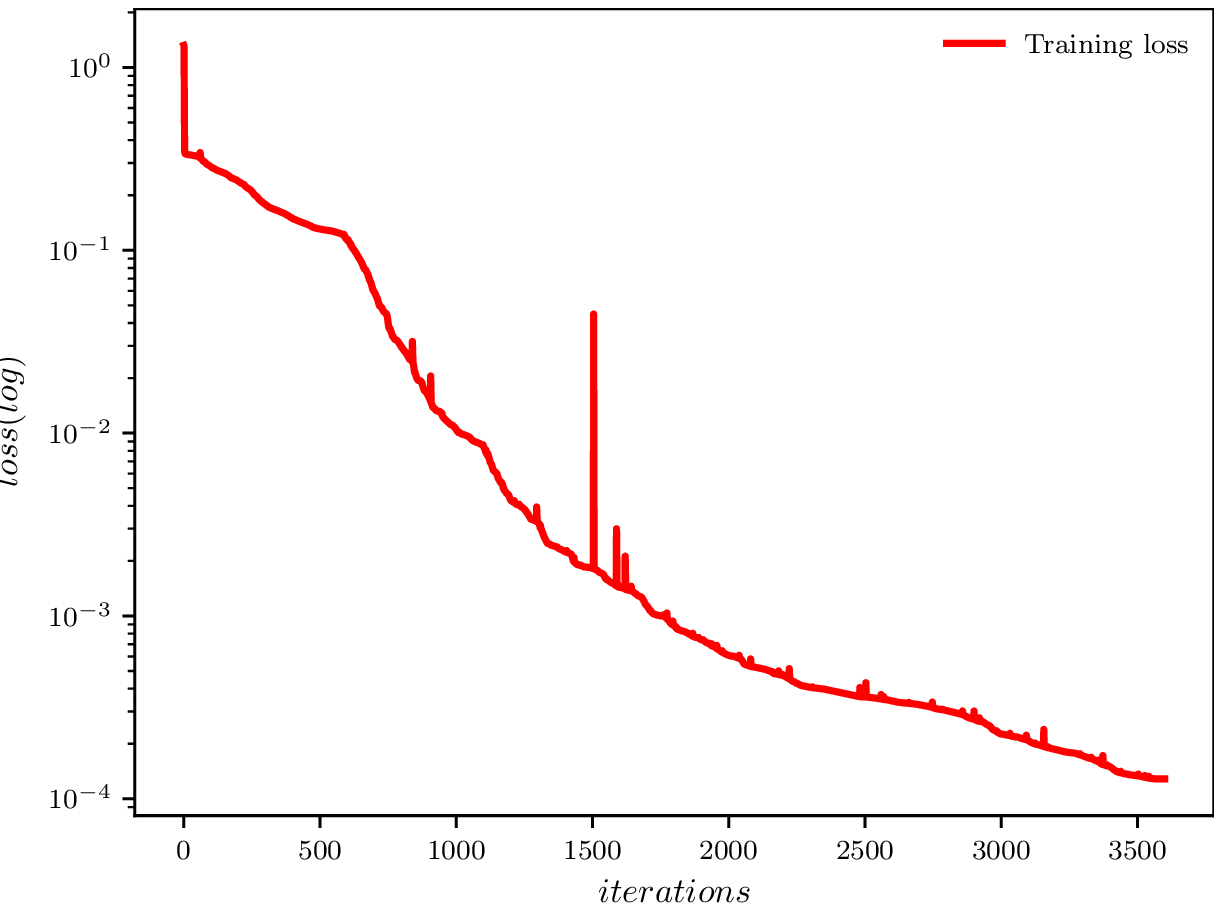}
$b$
\caption{The first order genuine rational soliton solution $q(x,t)$ based on the PINN: (a) The density plots and the sectional drawing; (b) The loss curve figure.}
\end{figure}

\begin{figure}
\centering
\includegraphics[width=6.5cm,height=5cm]{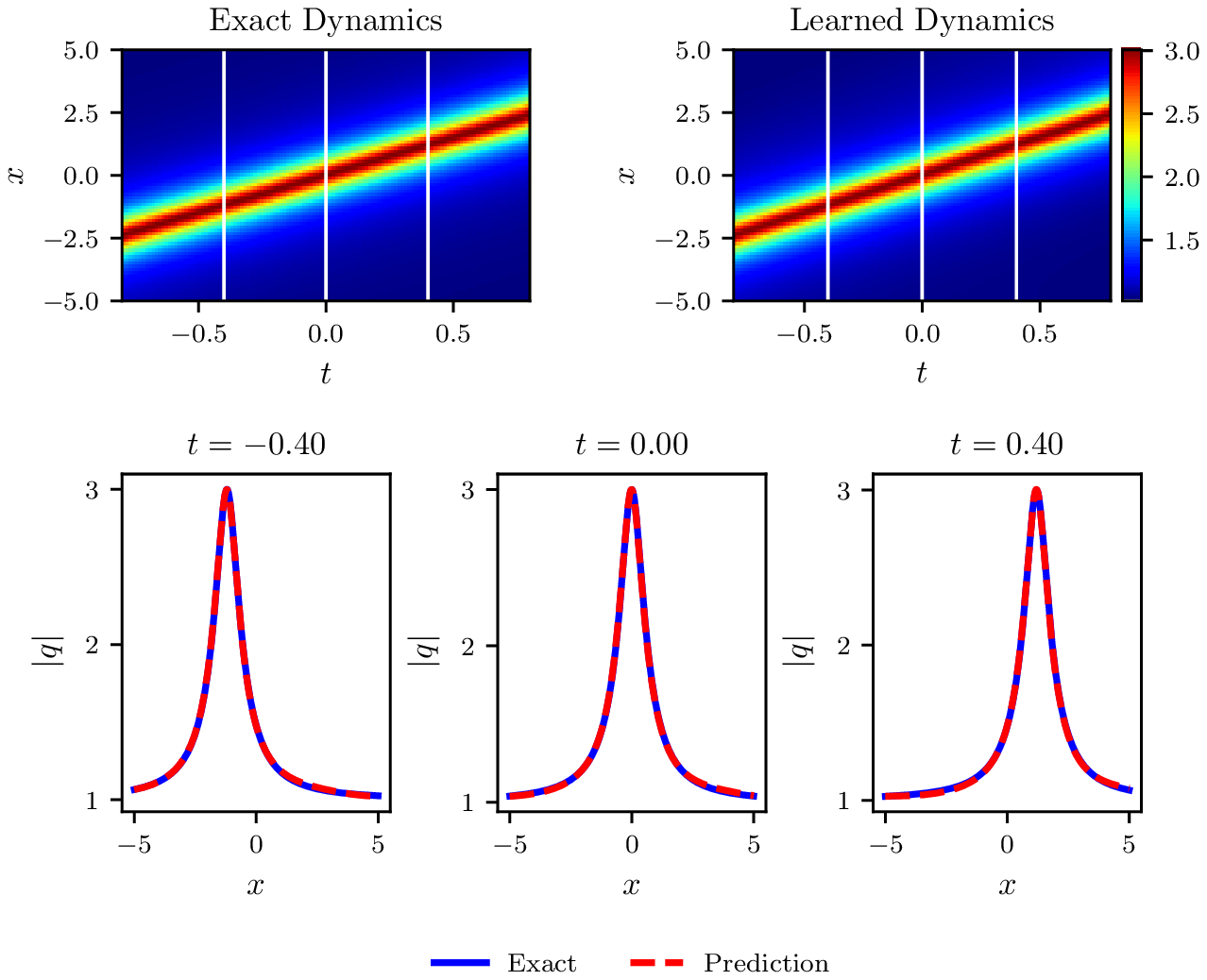}
$a$
\includegraphics[width=6.5cm,height=5cm]{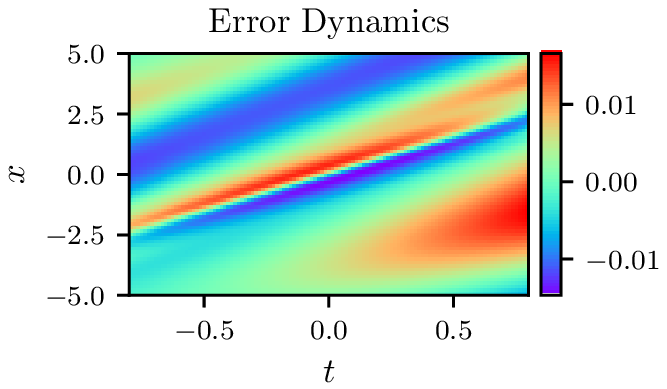}
$b$\\
\includegraphics[width=6.5cm,height=5cm]{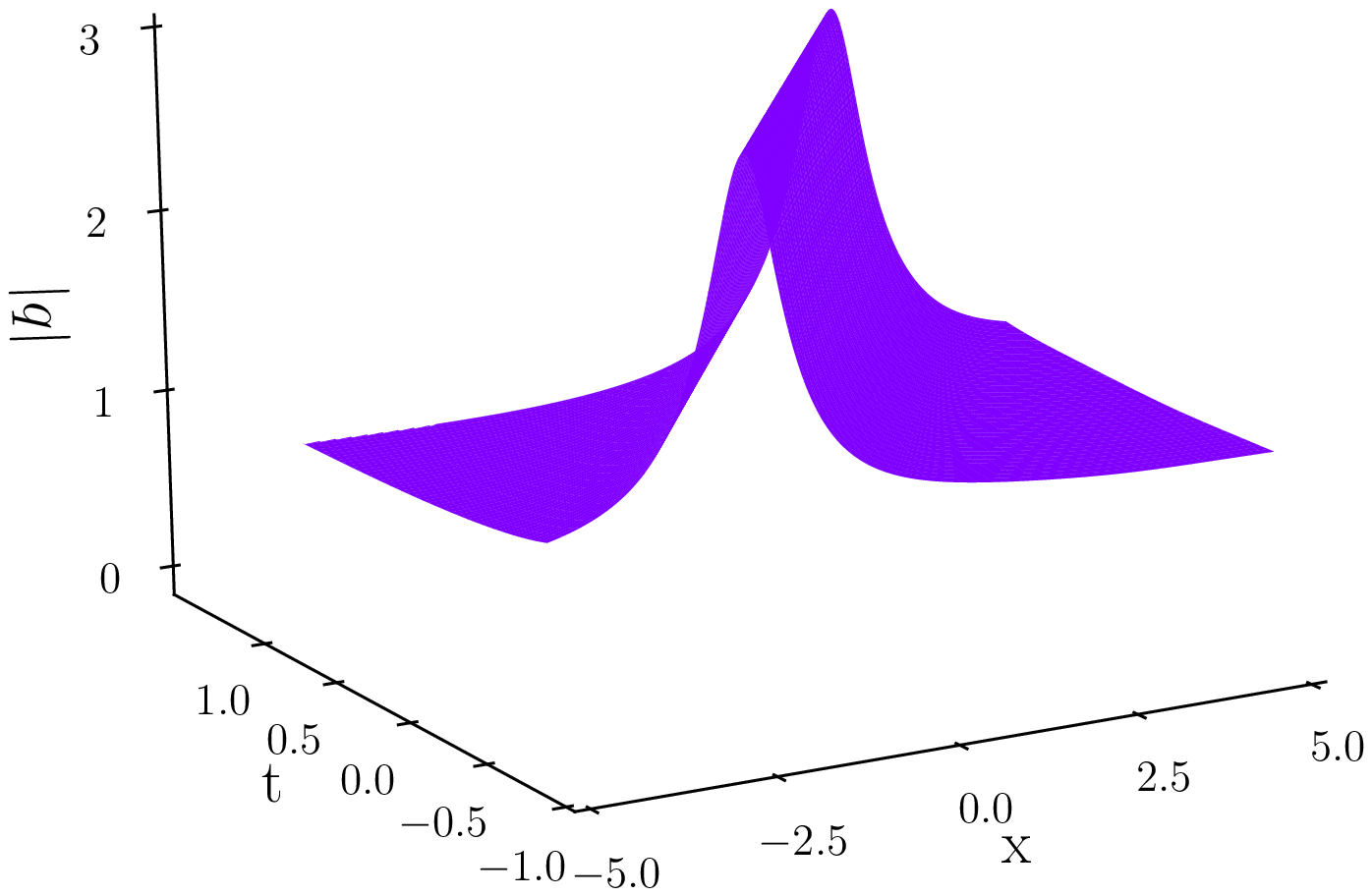}
$c$
\includegraphics[width=6.5cm,height=5cm]{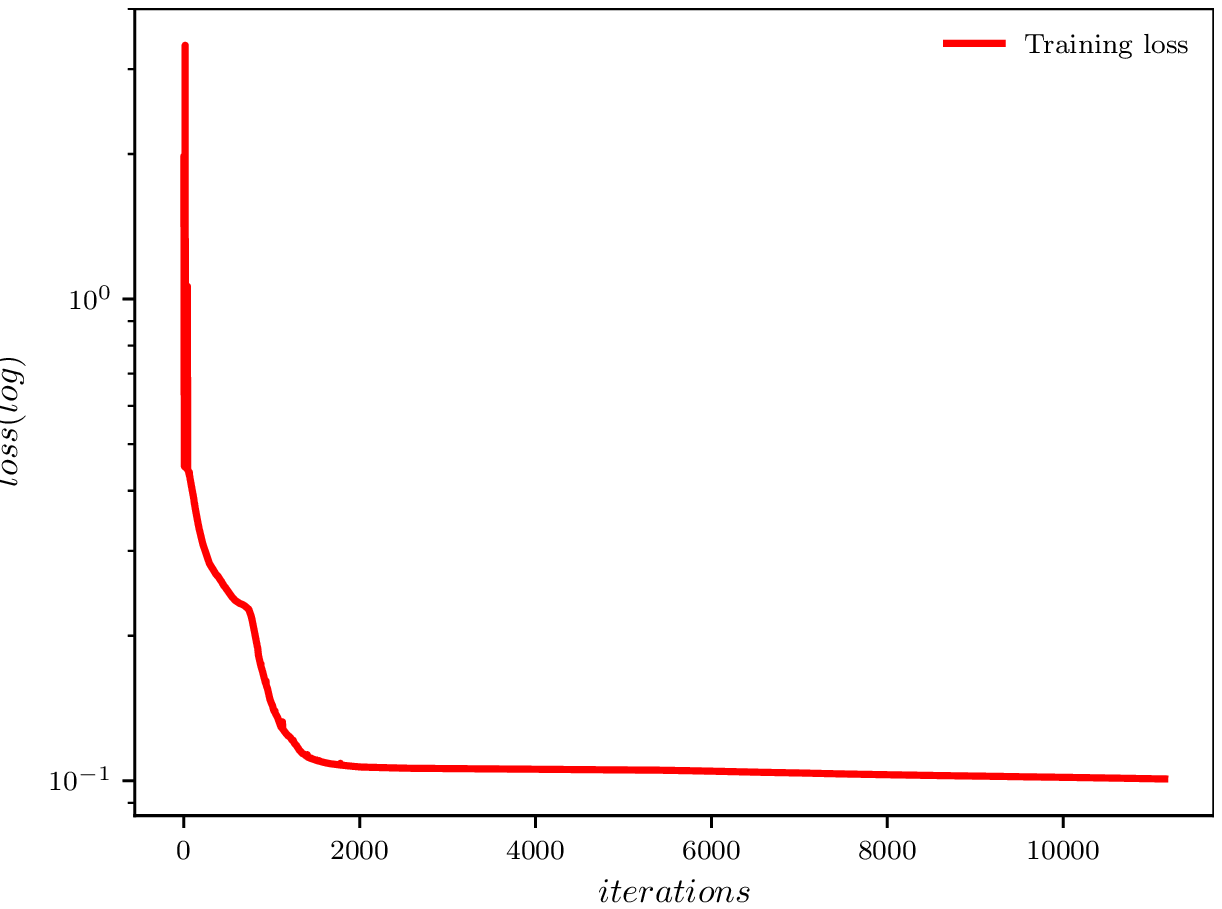}
$d$
\caption{The first order genuine rational soliton solution $q(x,t)$ based on the IPINN: (a) The density diagram and profiles at three different instants; (b) The error density diagram; (c) The three-dimensional motion; (d) The loss curve figure.}
\end{figure}

\section{Second order genuine rational soliton solution and two-order rogue wave solution of the DNLS}
In this section, we will use two diverse methods described in Section 2, which are consisted of PINN and IPINN, to construct the second order genuine rational soliton solution and two-order rogue wave solution of the DNLS, respectively. The detailed results and analysis are given out in the following two parts.

\subsection{Second order genuine rational soliton solution}
\quad

In this section, based on the Dirichlet boundary conditions Eq. \eqref{e15}, we will numerically predict the second order genuine rational soliton solution of the DNLS by using the PINN method and improved PINN method, separately. The second order genuine rational soliton solution of the DNLS has been derived in Ref. \cite{Guo2012}, the form is as follows
\begin{equation}\label{e21}
q(x,t)=\frac{L_1^*L_2}{L_1^2},
\end{equation}
where
\begin{equation}\nonumber
\begin{split}
&L_1=8(3t-x)^3+18(3t-x)+48t+12k+i[12(3t-x)^2+3],\\
&L_2=8(3t-x)^3-30(3t-x)+48t+12k+i[36(3t-x)^2-15],
\end{split}
\end{equation}
and $``*"$ denotes self conjugate, $k$ is an arbitrary real number. The norm of solution \eqref{e21} attains the maximum value five which locates at $(x,t)=\left(-\frac34k,-\frac14k\right)$, and vanishes Eq. \eqref{e21} at $\left(\frac{7\alpha-4\alpha^3-6k}{8},\frac{15\alpha-4\alpha^3-6k}{24}\right)$, where $\alpha=\pm\sqrt{\frac{5}{12}}$. The ``ridge" of this soliton \eqref{e21} approximately lays on the line $x=3t$. When $t\rightarrow\pm\infty$, above the second order genuine rational soliton solution \eqref{e21} approaches to the first order genuine rational soliton solution represented by \eqref{e17} along its ``ridge".

Then we take $[x_0,x_1]$ and $[t_0,t_1]$ in Eq. \eqref{e15} as $[-3.0,3.0]$ and $[-0.4,0.4]$, respectively. The corresponding initial condition is obtained by substituting $k=\mathrm{exp}(1)$ and the specific initial value into \eqref{e21}, we have
\begin{equation}\label{e22}
q_0(x)=\frac{L_1'^*L_2'}{L_1'^2},
\end{equation}
where
\begin{equation}\nonumber
\begin{split}
&L'_1=8(-1.2-x)^3+18(-1.2-x)-19.2+12\mathrm{exp}(1)+i[12(-1.2-x)^2+3],\\
&L'_2=8(-1.2-x)^3-30(-1.2-x)-19.2+12\mathrm{exp}(1)+i[36(-1.2-x)^2-15].
\end{split}
\end{equation}

Next, we obtain the initial and boundary value data set by the same data discretization method in Section 3.1. By randomly subsampling $N_q = 200$ from the original dataset and selecting $N_f = 20000$ configuration points, a training dataset composed of initial-boundary data and collocation points is generated with the help of LHS. Then the data set is substituted into two neural network models which composed of two different neural network algorithms to simulate the second order genuine rational soliton solution. After training, the neural network model of PINN achieves a relative $\mathbb{L}_2$ error of 3.680510$\mathrm{e-}$02 in about 705.3579 seconds, and the number of iterations is 6167. However, the network structure of IPINN achieves a relative $\mathbb{L}_2$ error of 4.295123$\mathrm{e-}$02 in about 874.1350 seconds, and the number of iterations is 6142.

The PINN experiment results have been summarized in Fig. 5, and we simulate the solution of $q(x,t)$ and obtain the density plots, profile, iterative curve plots of the second order genuine rational soliton solution. From (b) of Figure 5, it can be clearly observed that the curve of loss function declines very slowly, and there have a particularly large fluctuation after 6500 iterations, which indicate that the PINN has slow convergence and poor stability of loss function. Fig. 6 displays the training outcome by choosing the improved PINN method, and the density diagrams, profiles at different instants, error dynamics diagrams, three dimensional motion and loss curve figure of the second order genuine rational soliton solution $q(x,t)$ are illustrated. The top panel of (a) of Fig. 6 gives the density map of hidden solution $q(x,t)$, and when combing (b) of Fig. 6 with the bottom panel of (a) in Fig. 6, we can see that the relative $\mathbb{L}_2$ error is relatively large at $t\geqslant0.20$. From (d) of Fig. 6, in contrast with the first order genuine rational soliton solution by utilizing the improved PINN method in Section 3.2, the loss function curve of the second order genuine rational soliton solution is relatively stable, and the whole iterative process is relatively long, which is completely different from the sharp drop of the loss function curve about the first order genuine rational soliton solution and the less number of effective iterations in (d) of Fig. 4. In a word, from the two neural network methods, the results show that both the PINN and IPINN can simulate the second order genuine rational soliton solution accurately, and the training time, relative error and iteration number are similar, but the iterative process of IPINN is more stable and the training performance is better. There is no doubt that the IPINN is more reliable in training higher order solutions of the DNLS.

\begin{figure}
\centering
\includegraphics[width=6.5cm,height=5cm]{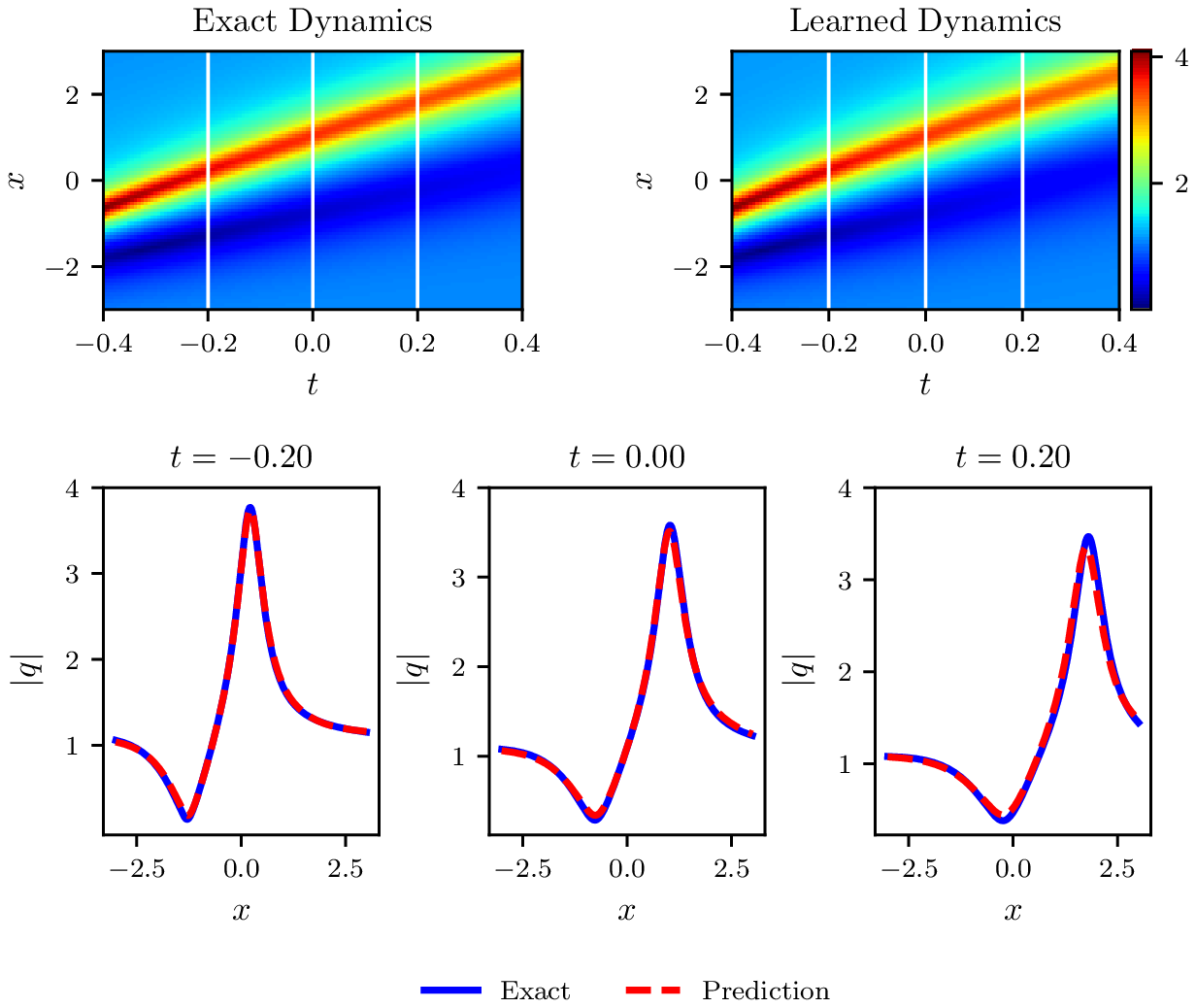}
$a$
\includegraphics[width=6.5cm,height=5cm]{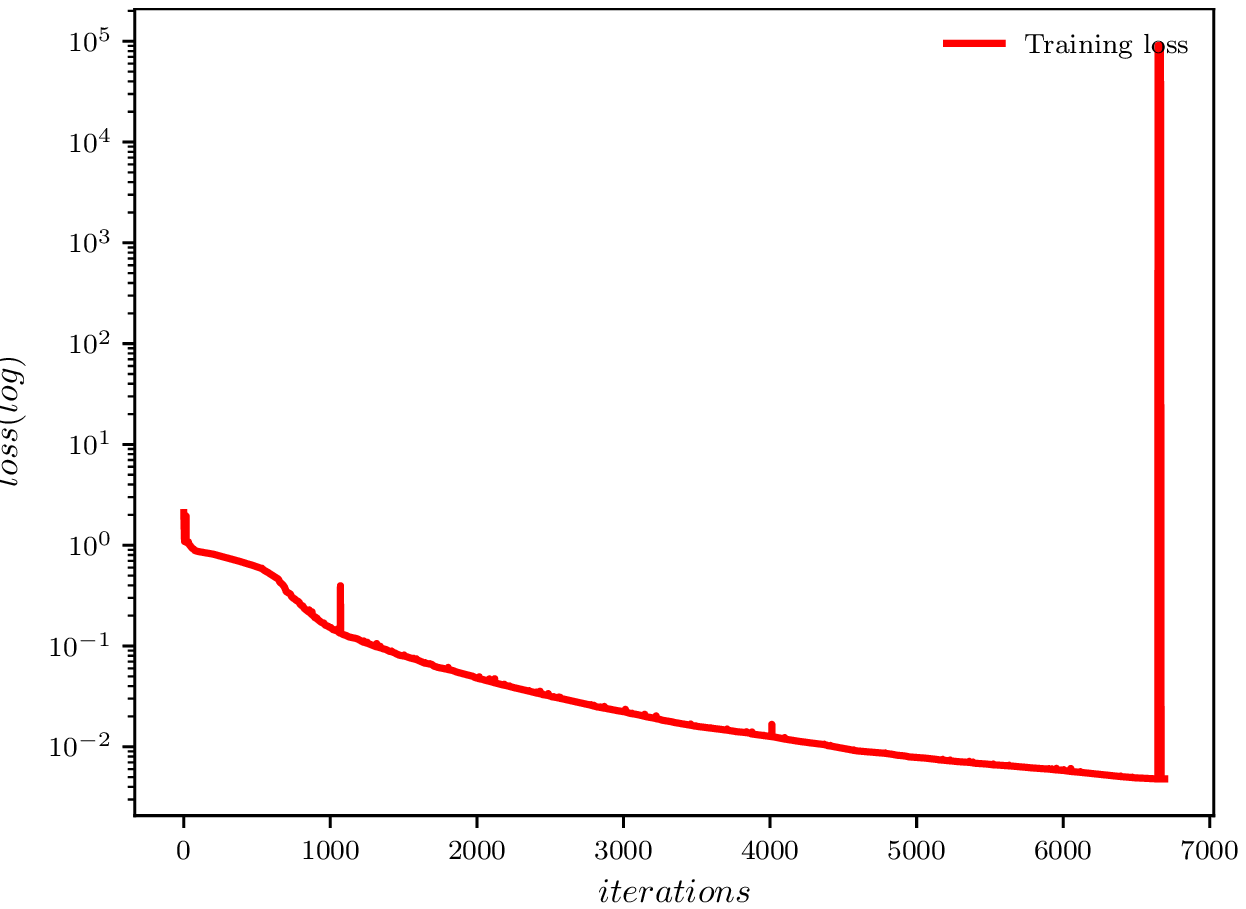}
$b$
\caption{The second order genuine rational soliton solution $q(x,t)$ based on PINN: (a) The density plots and the sectional drawing; (b) The loss curve figure.}
\end{figure}

\begin{figure}
\centering
\includegraphics[width=6.5cm,height=5cm]{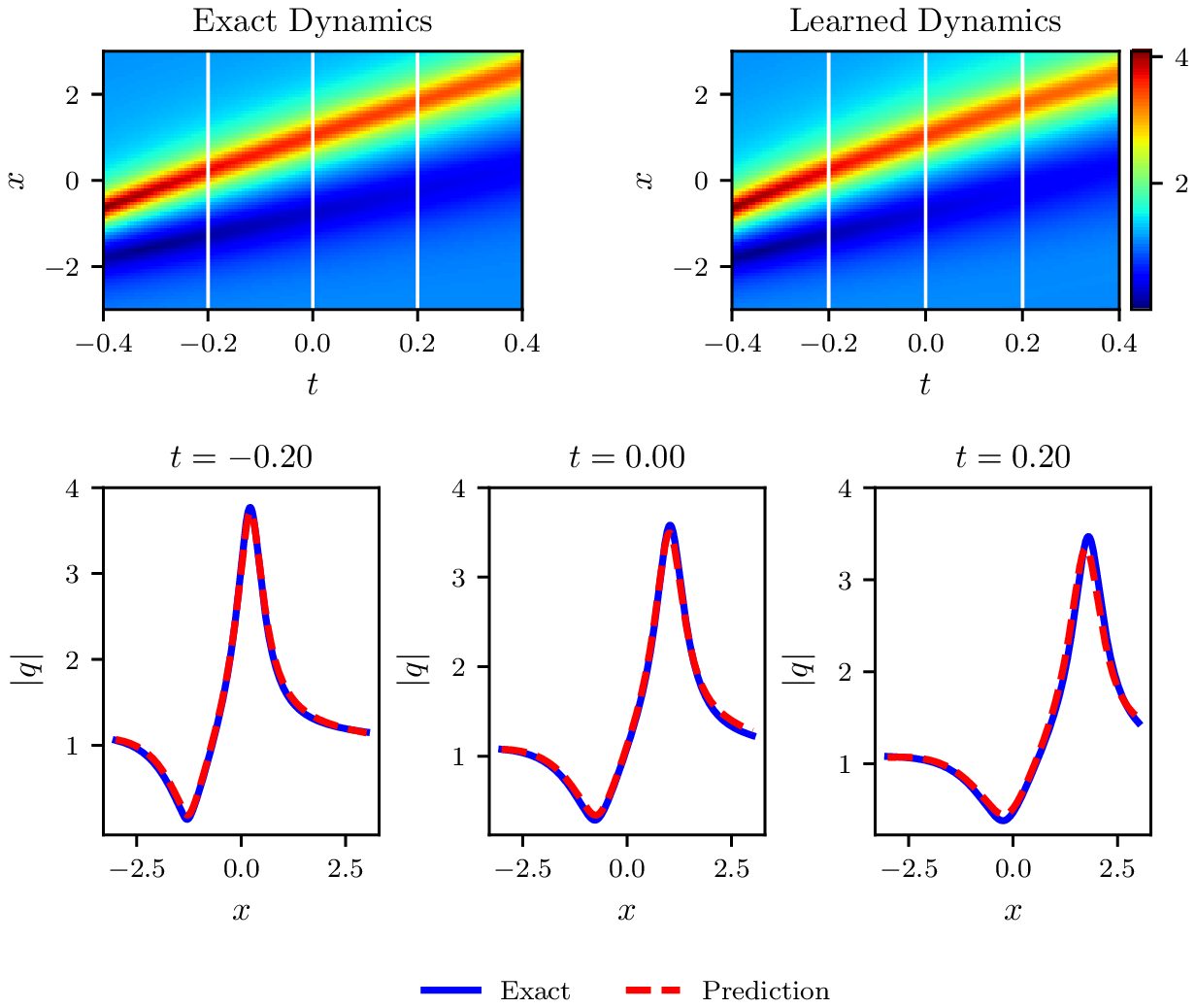}
$a$
\includegraphics[width=6.5cm,height=5cm]{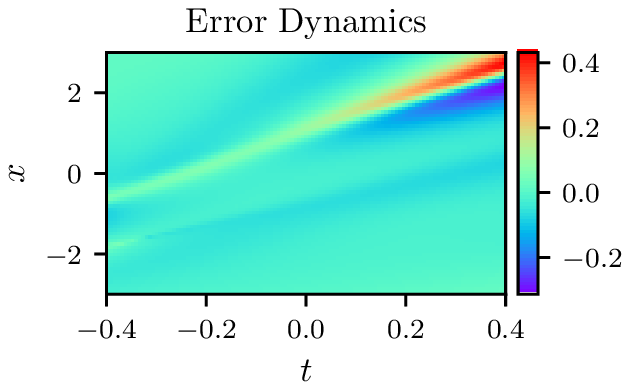}
$b$\\
\includegraphics[width=6.5cm,height=5cm]{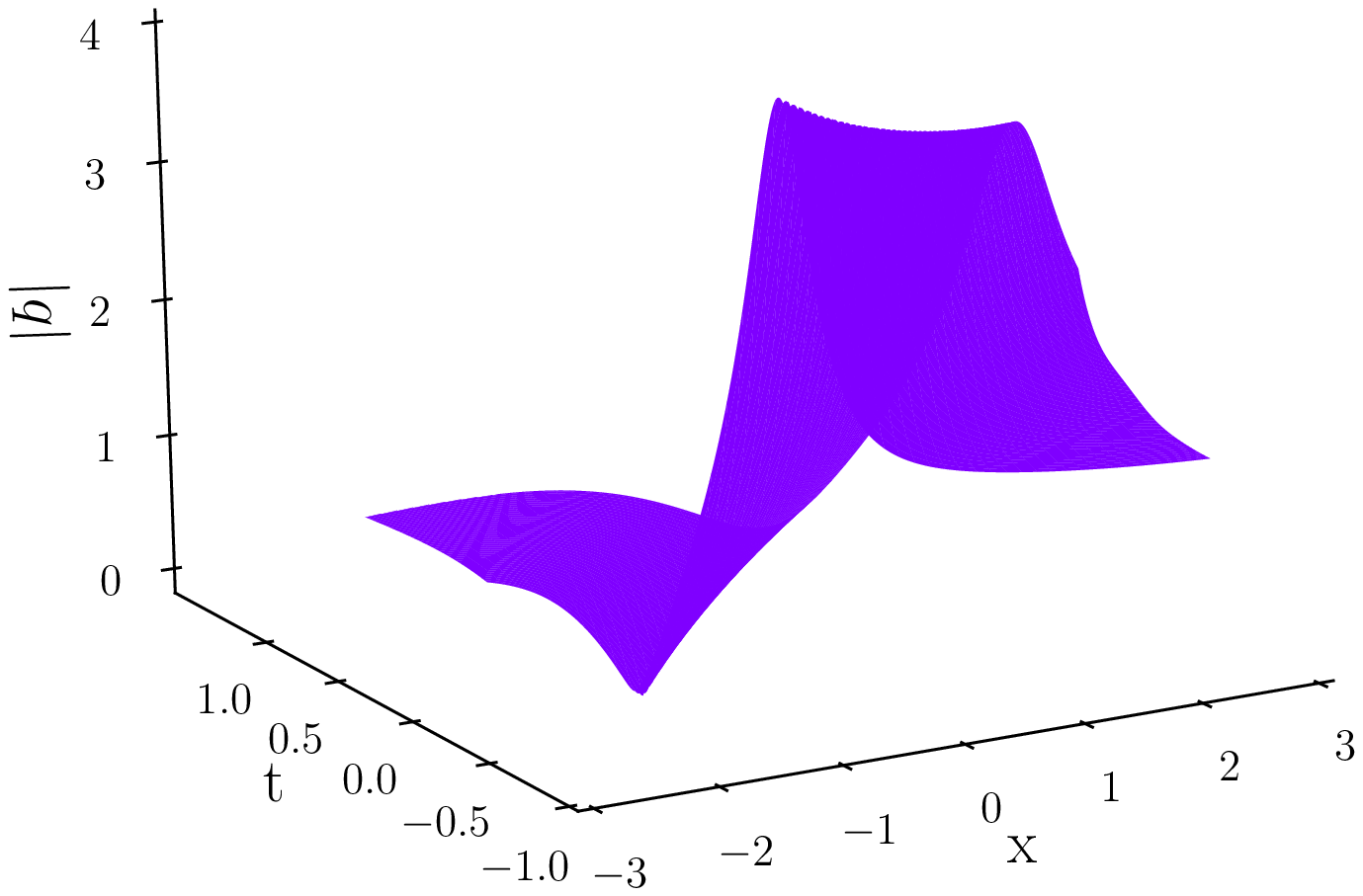}
$c$
\includegraphics[width=6.5cm,height=5cm]{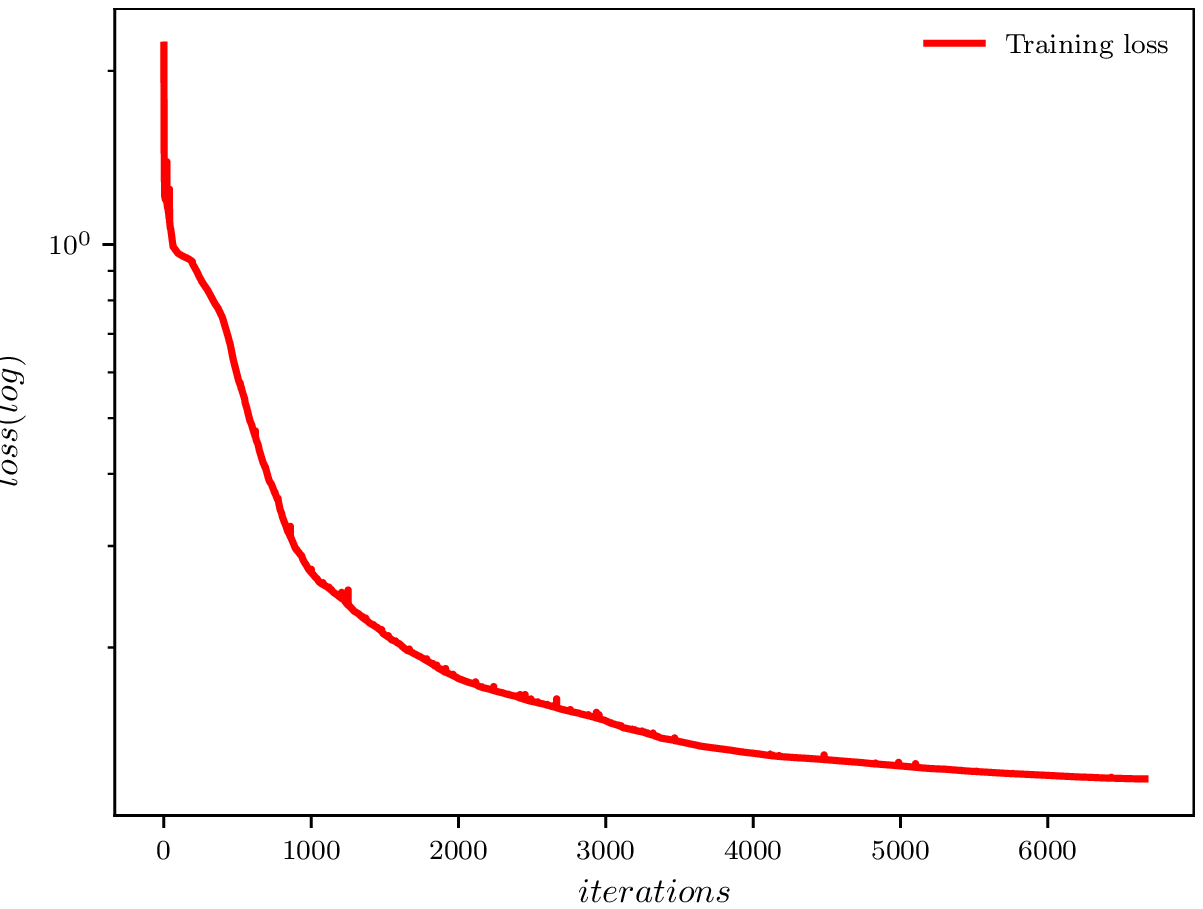}
$d$
\caption{The second order genuine rational soliton solution $q(x,t)$ based on IPINN: (a) The density plots and the sectional drawing; (b) The error density plots; (c) The three-dimensional plots; (d) The iterative curve plots.}
\end{figure}

In addition, according to the neural network model of IPINN, we obtain the following two tables specifically. Based on the same initial and boundary values of the second order genuine rational soliton solution in the case of $N_q=200$ and $N_f=20000$, we employ the control variable method which is often used in physic to study the effects of different levels of neural networks and different numbers of single-layer neurons on the second order genuine rational soliton solution dynamics of the DNLS. Moreover, the relative $\mathbb{L}_2$ error of different layers of neural networks and different numbers of single-layer neurons are given in Table 1. From the data in Table 1, we can see that when the number of neural network layers is fixed, the more the number of single-layer neurons, the smaller the relative $\mathbb{L}_2$ error. Of course, due to the influence of randomness, there are individual data results that do not meet the previous conclusion, but on the whole the conclusion is tenable. Similarly, when the number of single-layer neurons is fixed, the deeper the layer is, the smaller the relative error is. To sum up, we can draw the conclusion that the number of layers of neural network and the number of single-layer neurons jointly determine the relative $\mathbb{L}_2$ error, and when the number of layers is not less than 6 and the number of neurons in a single layer is not less than 30, the overall relative error is small. In the case of the same original dataset, Table 2 shows the relative $\mathbb{L}_2$ error of nine-layer neural network and single-layer neural network with 40 neurons when taking different number of sampling points $N_q$ in the initial-boundary data and different number of collocation points $N_f$ which are generated by the Latin hypercube sampling method. From the table 2, we can see that the influences of $N_q$ and $N_f$ on the relative $\mathbb{L}_2$ error of neural network are not so obvious. After careful observation, when taking $N_f=20000$, regardless of the number of $N_q$, the overall relative $\mathbb{L}_2$ error is small, which also explain why the neural network model can simulate more accurate numerical solutions with smaller initial data set.

\begin{table}[htbp]
  \caption{The second order genuine rational soliton solution of the DNLS by using the IPINN: Relative final prediction error measure in the $\mathbb{L}_2$ norm for different number of hidden layers and neurons in each layer.}
  \centering
  \begin{tabular}{p{3.38cm}|p{2cm}p{2cm}p{2cm}p{2cm}p{2cm}}
  \toprule
  \textbf{\diagbox{Layers}{Neurons}} &\textbf{\quad\quad20} &\textbf{\quad\quad30} &\textbf{\quad\quad40} &\textbf{\quad\quad50} &\textbf{\quad\quad60}\\
  \midrule
  \textbf{3}   &5.745666e-01&5.788318e-01&4.960168e-01&1.596054e-01&4.453853e-01\\
  \textbf{6}   &5.945028e-01&5.162760e-02&6.452197e-02&1.540266e-01&7.040157e-02\\
  \textbf{9}   &2.476185e-01&1.089718e-01&4.295123e-02&1.045272e-01&1.788330e-02\\
  \textbf{12}  &3.268944e-01&5.060934e-02&6.087790e-02&5.869449e-02&1.037406e-01\\
  \bottomrule
  \end{tabular}
\end{table}

\begin{table}[htbp]
  \caption{The second order genuine rational soliton solution of the DNLS by using the IPINN: Relative final prediction error measure in the $\mathbb{L}_2$ norm for different number of $N_q$ and $N_f$.}
  \label{Tab:bookRWCal}
  \centering
  \begin{tabular}{p{1.35cm}|p{2cm}p{2cm}p{2cm}p{2cm}p{2cm}}
  \toprule
  \textbf{\diagbox{$N_q$}{$N_f$}} &\textbf{\quad16000} &\textbf{\quad18000} &\textbf{\quad20000} &\textbf{\quad22000} &\textbf{\quad24000}\\
  \midrule
  \textbf{150}   &6.392143e-01&2.899009e-01&1.041697e-01&6.879006e-01&5.652572e-01\\
  \textbf{200}   &6.728854e-01&5.891113e-01&4.295123e-02&1.944170e-01&6.393751e-01\\
  \textbf{250}   &5.824423e-01&3.160742e-02&2.692156e-02&6.872238e-01&6.428133e-01\\
  \bottomrule
  \end{tabular}
\end{table}

\subsection{Two-order rogue wave solution}
\quad

Recently, the study of rogue waves is one of the hot topics in many areas including optics, plasma, ocean dynamics, machine learning, Bose-Einstein condensate and even finance and so on\cite{PuJ2020,Solli2007,Yue2020,Marcucci2020,WangM2019,Yan2011,ZhangX2019}. In addition to the peak amplitude more than twice of the background wave, rogue waves also have the characteristics of instability and unpredictability. Therefore, the researches and applications of rogue waves play an momentous role in real life, especially how to avoid the damage to ships caused by ocean rogue waves is of great practical significance. At present, Marcucci et al. have investigated the computational machine in which nonlinear waves replace the internal layers of neural networks, discussed learning conditions, and demonstrated functional interpolation, data interpolation, data sets, and Boolean operations. When the nonlinear Schr\"odinger equation is considered, the use of highly nonlinear regions means that solitons, rogue waves and shock waves play a leading role in the training and calculation \cite{Marcucci2020}. Moreover, the dynamical behaviors and error analysis about the one-order and two-order rogue waves of the nonlinear Schr\"{o}dinger equation have been revealed by the deep learning neural network with physical constraints for the first time \cite{PuJ2020}. The rogue wave solutions of the DNLS were derived in via Darboux transformation \cite{XuS2011}, and the high-order rogue wave solutions are obtained by generalized Darboux transformation \cite{Guo2012}. However, to the best of our knowledge, the machine learning with neural network model has not been exploited to simulate the rogue wave solution of the DNLS. In this section, we construct the two-order rogue wave solution of the DNLS by employing the PINN and IPINN, respectively. Some vital comparisons are given out to better describe the advantages of PINN and IPINN.

On the basis of the Dirichlet boundary conditions Eq. \eqref{e15}, we will numerically training the two-order rogue wave solution of the DNLS by employing the PINN method and improved PINN method, separately. The two-order rogue wave solution of the DNLS has been derived in Ref. \cite{Guo2012}, the form can be represented as following
\begin{equation}\label{e23}
q(x,t)=\frac{R_1^*R_2}{R_1^2}\mathrm{exp}(-ix),
\end{equation}
where
\begin{equation}\nonumber
\begin{split}
&R_1=8x^6+24x^4t^2+24x^2t^4+8t^6+24ix^5-24ix^4t+48ix^3t^2-48ix^2t^3+24ixt^4-24it^5-12x^4+48x^3t-216x^2t^2\\
&\quad\quad+48xt^3+180t^4+48ix^3-288ixt^2-336it^3+90x^2-72xt+666t^2+54ix-198it+9,\\
&R_2=8x^6+24x^4t^2+24x^2t^4+8t^6-24ix^5-72ix^4t-48ix^3t^2-144ix^2t^3-24ixt^4-72it^5-60x^4-144x^3t-504x^2t^2\\
&\quad\quad-144xt^3-60t^4+48ix^3+288ix^2t+576ixt^2-528it^3-198x^2+504xt-486t^2+90ix+414it+45.
\end{split}
\end{equation}

Then we take $[x_0,x_1]$ and $[t_0,t_1]$ in Eq. \eqref{e15} as $[-2.5,2.5]$ and $[-0.01,0.01]$, respectively. The corresponding initial condition is obtained after substituting the specific initial value into \eqref{e23}, we have
\begin{equation}\label{e24}
q_0(x)=\frac{R_1'^*R_2'}{R_1'^2}\mathrm{exp}(-ix),
\end{equation}
where
\begin{equation}\nonumber
\begin{split}
&R'_1=8x^6-11.9976x^4-0.48x^3+89.97840024x^2+0.719952x+9.0666018+\\
&\quad\quad i(24x^5+0.24x^4+48.0048x^3+0.000048x^2+53.97120024x+1.980336002),\\
&R'_2=8x^6-59.9976x^4+1.44x^3-198.0503998x^2-5.039856x+44.9513994+\\
&\quad\quad i(-24x^5+0.72x^4+47.9952x^3-2.879856x^2+90.05759976x-4.139471993).
\end{split}
\end{equation}

Similar to the discretization method in Section 3.1, we randomly sample $N_q=300$ from the original initial boundary value condition data set, and select $N_f=20000$ configuration points which are generated by the LHS method. Thus, the training dataset of initial boundary value data and configuration points are formed. After training with two methods, the neural network model of PINN achieves a relative $\mathbb{L}_2$ error of 8.412217e-02 in about 1188.4475 seconds, and the number of iterations is 9470.  Moreover, introducing the IPINN, the structure attains a relative $\mathbb{L}_2$ error of 7.262528e-02 in about 2924.0589 seconds, and the number of iterations is 18394. It can be seen from the above results that under the same experimental conditions, the relative $\mathbb{L}_2$ error of IPINN method is smaller than that of PINN for simulating rogue wave solution, but the improved PINN method has longer training time and more iterations. Next, we will give the specific numerical results and correlation analysis.

The density plots, the sectional drawing and the error density plots of the two-order rogue wave solution are exhibited by employing the PINN method in Fig. 7. In bottom panel of (a) in Figure 7, one can observe that the wave peak of the two-order rogue wave is well simulated, but the simulation on both sides of the wave peak is poor, which can also be verified from the error diagram in figure (b) in Fig. 7. On the other hand, as for the neural network model which applying the improved PINN method, its density plot, section drawing, error density plots, three-dimensional diagram and loss function curve diagram are shown in detail in Figure 8. Similarly, we find that the wave peak simulation in (a) of Fig. 8 is not as good as that in Figure 7, but the simulation on both sides of the wave peak is better, which is just opposite to the situation simulated of the PINN. On the whole, the simulation satisfaction of IPINN is higher, and it has more research value. The chart (b) in Figure 8 shows that there is a little error at the middle peak, where the error is the difference value between the accurate solution and the predicted solution. The 3D plots and the loss function curve are shown in (c) and (d) of Figure 8, respectively. From figure (d) of Fig. 8, it can be seen that the loss value fluctuates greatly when the number of iterations is around 2500, and then decreases slowly from 1 to 0.1.

\begin{figure}
\centering
\includegraphics[width=6.5cm,height=5cm]{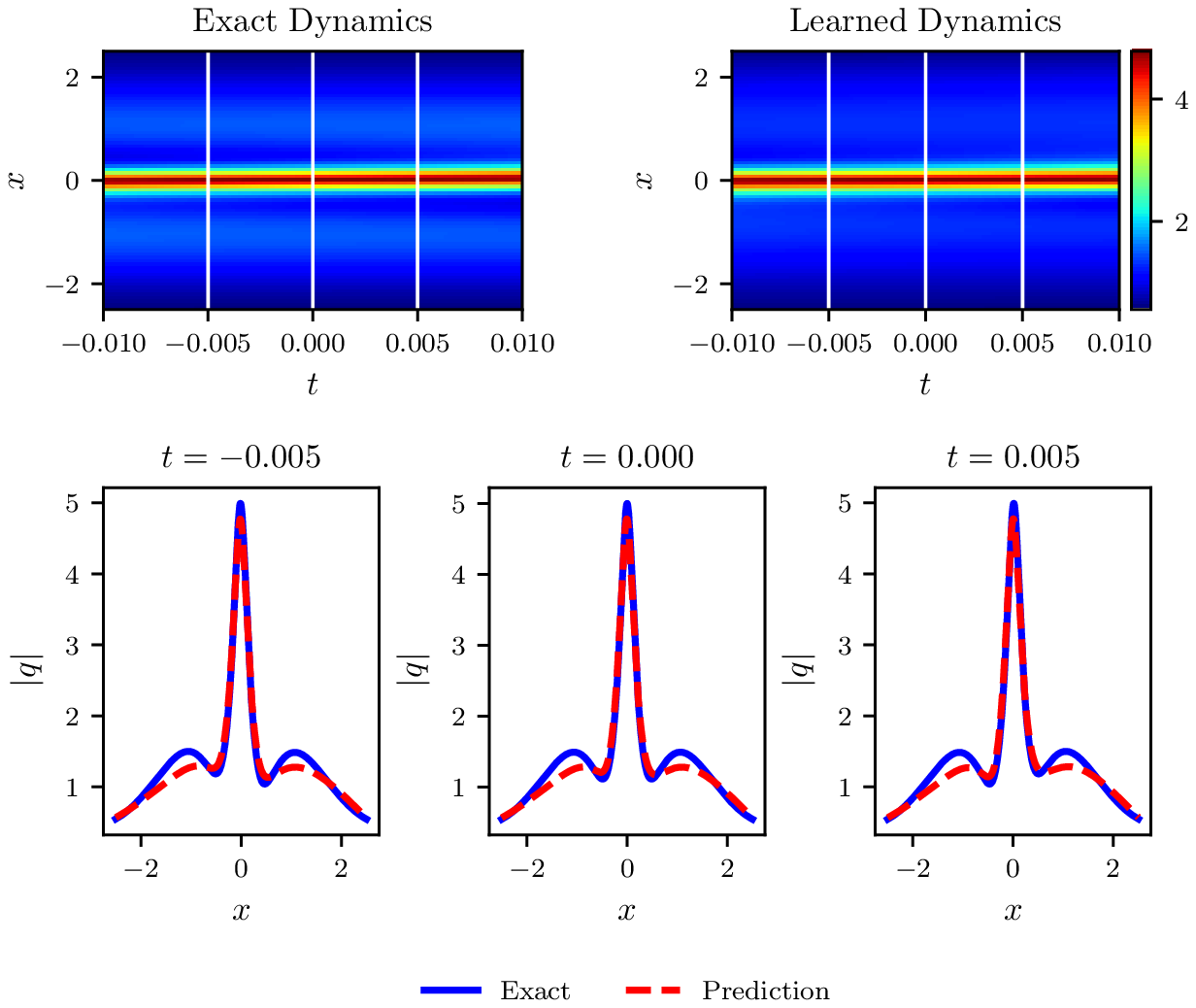}
$a$
\includegraphics[width=6.5cm,height=5cm]{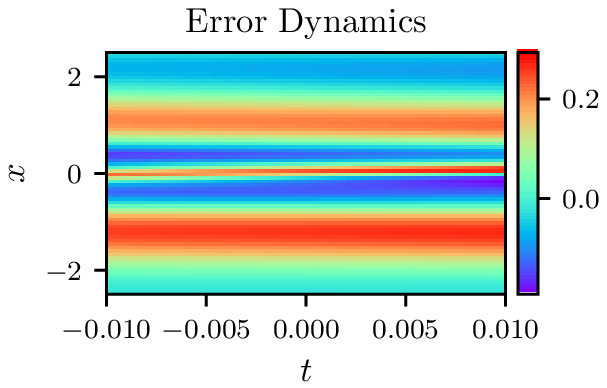}
$b$
\caption{The two-order rogue wave solution $q(x,t)$ based on the PINN: (a) The density plots and the sectional drawing; (b) The error density plots.}
\end{figure}

\begin{figure}
\centering
\includegraphics[width=6.5cm,height=5cm]{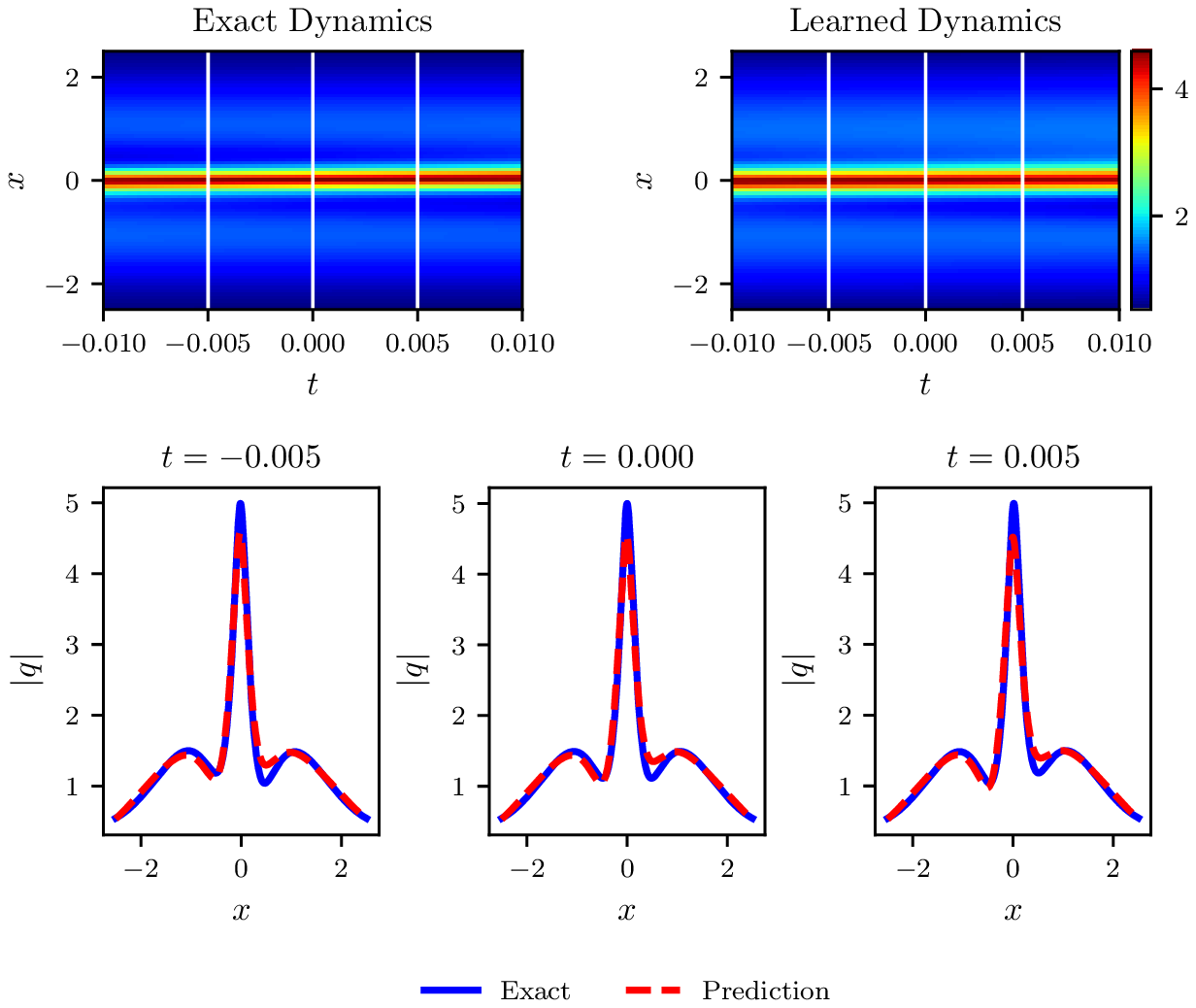}
$a$
\includegraphics[width=6.5cm,height=5cm]{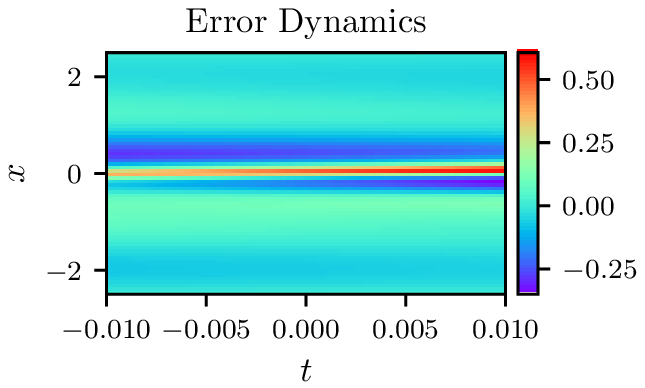}
$b$\\
\includegraphics[width=6.5cm,height=5cm]{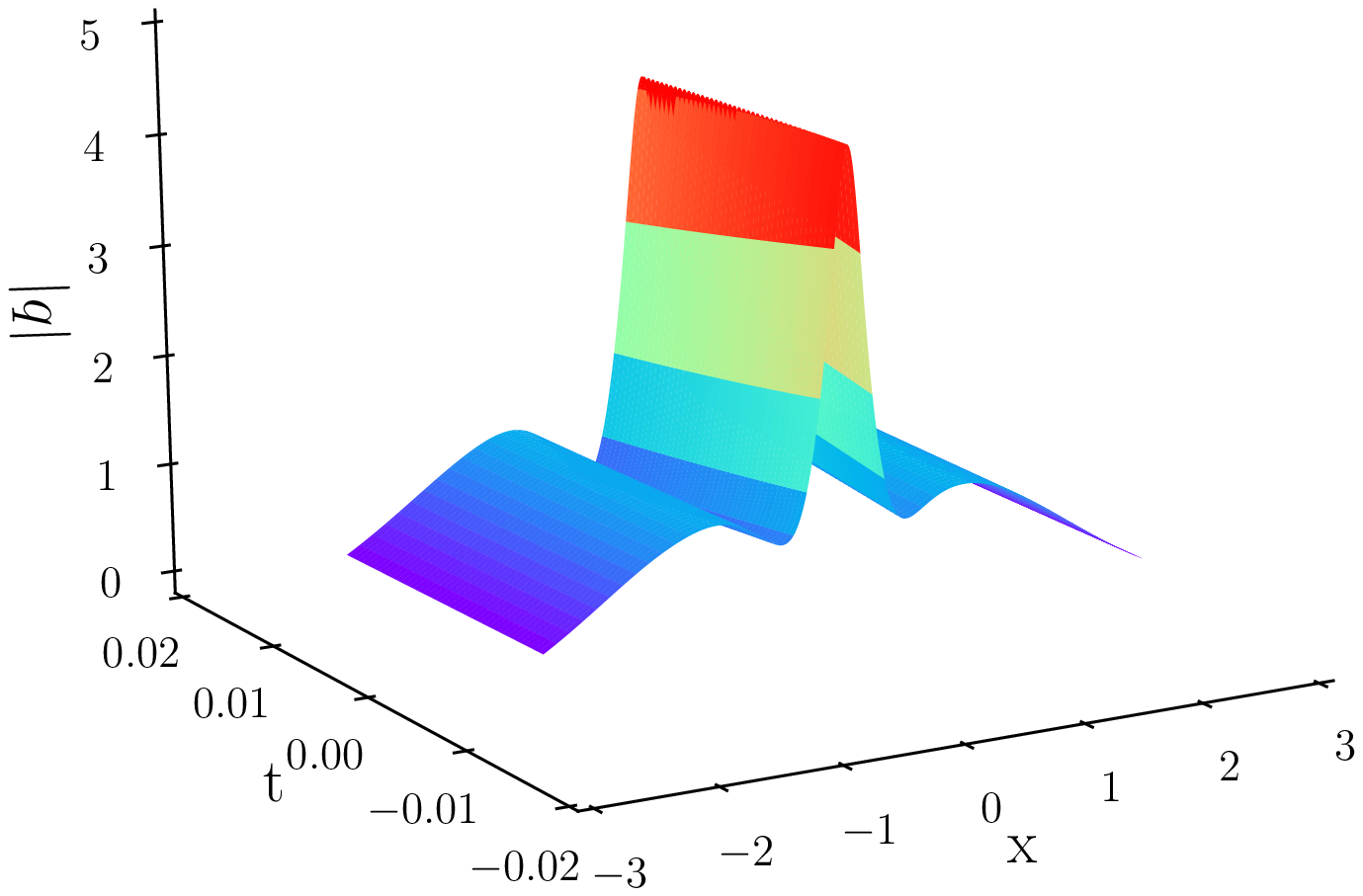}
$c$
\includegraphics[width=6.5cm,height=5cm]{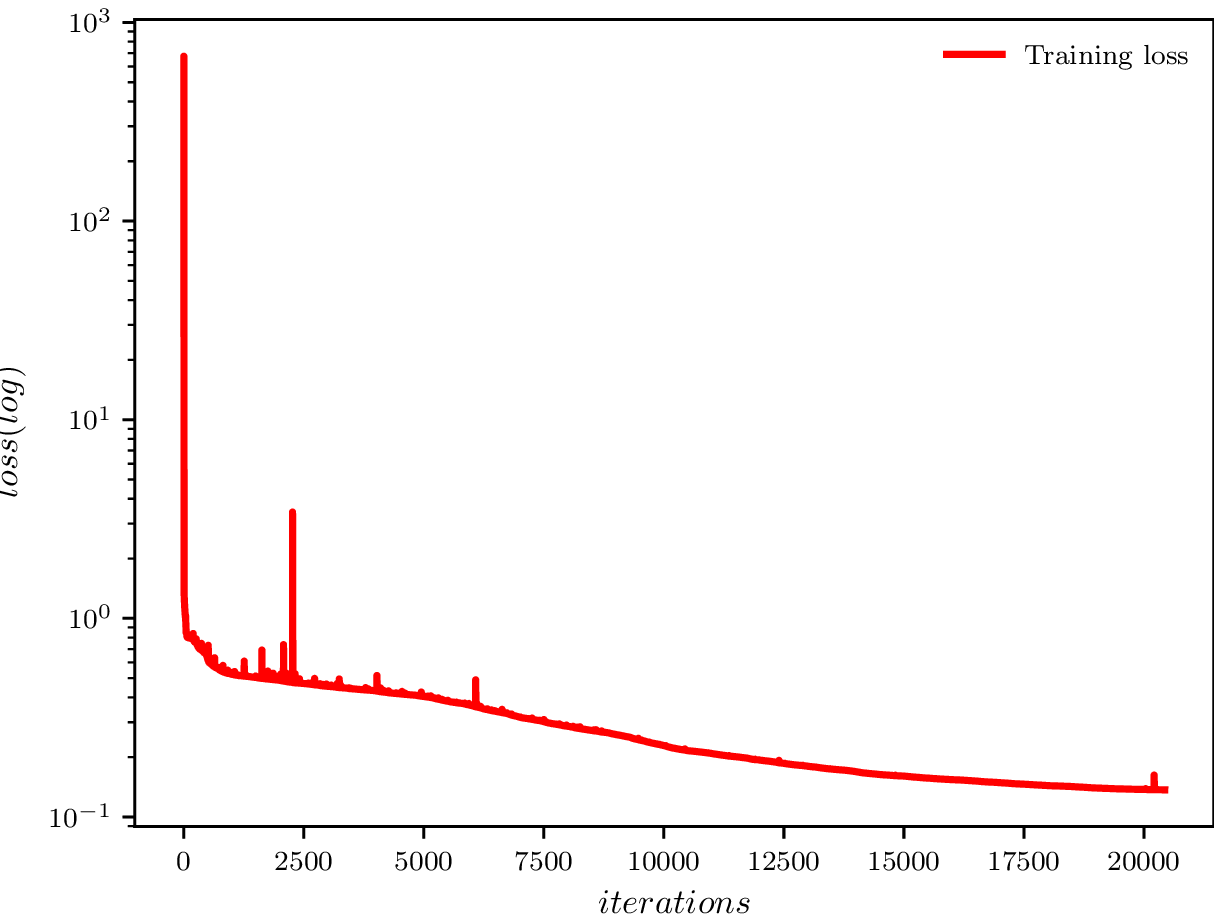}
$d$
\caption{The two-order rogue wave solution $q(x,t)$ based on the IPINN: (a) The density plots and the sectional drawing; (b) The error density plots; (c) The three-dimensional plots; (d) The iterative curve plots.}
\end{figure}

Futhermore, initialization of the scaled parameters can be done in various ways as long as such value does not cause divergence of the loss. In this work, the scaled parameters are initialized as $na_i^m=1,\forall n\geqslant1$. Although, an increase in scaling factor speeds up the convergence rate, at the same time the parameter $a_i^m$ becomes more sensitive. In order to better understand the influence of initialization of scalable parameters on the improved PINN algorithm model, we present four different initialization conditions of scalable parameters to obtain the two-order rogue wave solution by employing the improved PINN method in Table 3. From the Table 3, we can drastically observe that when amplify the scaled hyper-parameter $n$ in the initialization conditions of scalable parameters, the number of iterations and training time increase, but the relative $\mathbb{L}_2$ error does not blindly dwindle. When the hyper-parameter $n=10$, the relative $\mathbb{L}_2$ error is minimum and the training effect is better in Table 3. This also reveals why we generally choose the initialization of scalable parameters as $n=10,a_i^m=0.1$ in the IPINN with locally adaptive activation function in this paper.
\begin{table}[htbp]
  \caption{The two-order rogue wave solution of the DNLS by utilizing the IPINN: Relative $\mathbb{L}_2$ norm error, training time and iterations for different initialization conditions of scalable parameters(ICSP).}
  \label{Tab:bookRWCal}
  \centering
  \begin{tabular}{p{2.32cm}|p{2.8cm}p{2.8cm}p{2.8cm}p{2.8cm}}
  \toprule
  \textbf{\diagbox{Type}{ICSP}} &Variable a,(n=1) &Variable a,(n=5) &Variable a,(n=10) &Variable a,(n=20)\\
  \midrule
  Relative error &\quad 1.309705e-01&\quad 8.760463e-02&\quad 7.262528e-02&\quad 1.191116e-01\\
  Training time  &\quad\quad 1499.7196&\quad\quad 2203.6984&\quad\quad 2924.0589&\quad\quad 3983.7843\\
  \quad Iterations  &\quad\quad\quad 9194&\quad\quad\quad 14385&\quad\quad\quad 18394&\quad\quad\quad 24043\\
  \bottomrule
  \end{tabular}
\end{table}

The rogue wave is a kind of wave that comes and goes without trace, the research of seeking and simulating solution can provide an significant theoretical basis for the prediction and utilization of the rogue waves. Compared with the nonlinear Schr\"odinger equation, the form of rogue wave solution of the DNLS is more complex. We have successfully utilized the PINN to simulate the rogue wave solutions of the nonlinear Schr\"odinger equation in Ref. \cite{PuJ2020}. In this section, a large number of experiments and analysis have been carried out, and finally the two-order rogue wave solution of the DNLS has been imitated. In term of the same experimental conditions and environment, the PINN is better at simulating the wave crest, and the IPINN has better comprehensive effect on wave crest and both sides of wave crest. Apparently, the IPINN has more advantages about the overall effect, especially in simulation of the more complex rogue wave solutions.

\section{Conclusion}
Compared with traditional numerical methods, the PINN method has no mesh size limits and gives full play to the advantages of computer science. Moreover, due to the physical constraints, the neural network is trained with remarkably few data and fast convergence rate, and has a better physical interpretability. These numerical methods showcase a series of results of various problems in the interdisciplinary field of applied mathematics and computational science which open a new path for using machine learning to simulate unknown solutions and correspondingly discover the parametric equations in scientific computing. It also provides a theoretical and practical basis for dealing with some high-dimensional scientific problems that can not be solved before.

In this paper, based on the PINN method, an improved PINN method which contains the locally adaptive activation function with scalable parameters is introduced to solve the classical integrable DNLS. The improved PINN method achieves a better performance of the neural network through such learnable parameters in the activation function. Specifically, applying two data-driven algorithms which including the PINN and IPINN to deduce the localized wave solutions which consist of the one-rational soliton, genuine rational soliton solutions and rogue wave solution for the DNLS. In all these cases, compared with the original PINN method, it is shown that the decay of loss function is faster in the case of the improved PINN method, and correspondingly the relative $\mathbb{L}_2$ error in the simulation of solution is shown to be similar or even smaller in the proposed approach. We outline how different types of localized wave solutions are generated due to different choices of initial and boundary value data. Remarkably, these numerical results show that the improved PINN method with locally adaptive activation function is more powerful than the PINN method in exactly recovering the different dynamic behaviors of the DNLS.

The improved PINN approach is a promising and powerful method to increase the efficiency, robustness and accuracy of the neural network based approximation of nonlinear functions as well as abundant localized wave solutions of integrable equations. Furthermore, more general nonlinear integrable equation, such as the Hirota equation which has been widely concerned in integrable systems, is not investigated in our work. Due to the ability of the improved PINN to accelerate the convergence rate and improve the network performance, more complex integrable equations could also be considered, such as the Kaup-Newell systems, Sasa-Satsuma equation, Camassa-Holm equation and so on. How to combine machine learning with integrable system theory more deeply and build significant integrable deep learning algorithm is an urgent problem to be solved in the future. These new problems and challenges will be considered in the future research.

\end{document}